\documentclass[twocolumn,prb,showpacs]{revtex4}%
\usepackage{graphicx}%
\usepackage{amsmath}%
\usepackage{amsfonts}%
\usepackage{amssymb}
\usepackage{bm}

\def\s{{\sigma}}
\def\e{{\epsilon}}
\def\k{{ {\bm k} }}

\def\q{{ {\bm q} }}
\def\Q{{ {\bm Q} }}

\def\x{{ {\bm x} }}
\def\w{{\omega}}
\def\a{{\alpha}}
\def\b{{\beta}}

\begin{document}
\title{
Origin of Orthorhombic Transition, Magnetic Transition,\\
and Shear Modulus Softening in Iron Pnictide Superconductors: \\
Analysis based on the Orbital Fluctuation Theory
}
\author{Hiroshi \textsc{Kontani}$^{1}$,
Tetsuro \textsc{Saito}$^{1}$, and
Seiichiro \textsc{Onari}$^{2}$
}
\date{\today }

\begin{abstract}
The main features in iron-pnictide superconductors are summarized as
(i) the orthorhombic transition
accompanied by remarkable softening of shear modulus, 
(ii) high-$T_{\rm c}$ superconductivity close to the orthorhombic phase, 
and (iii) stripe-type magnetic order induced by orthorhombicity.
To present a unified explanation for them,
we analyze the multiorbital Hubbard-Holstein model
with Fe-ion optical phonons based on the orbital fluctuation theory.
In the random-phase-approximation (RPA), 
a small electron-phonon coupling constant ($\lambda\sim0.2$)
is enough to produce large orbital (=charge quadrupole) fluctuations.
The most divergent susceptibility
is the $O_{xz}$-antiferro-quadrupole (AFQ) susceptibility, 
which causes the $s$-wave superconductivity without sign reversal 
($s_{++}$-wave state).
At the same time, divergent development of 
$O_{x^2-y^2}$-ferro-quadrupole (FQ) susceptibility
is brought by the ``two-orbiton process'' 
with respect to the AFQ fluctuations, which is absent in the RPA.
The derived FQ fluctuations cause the softening of 
$C_{66}$ shear modulus, and its long-range-order 
not only triggers the orthorhombic structure transition,
but also induces the instability of stripe-type antiferro-magnetic state.
In other words, the condensation of composite bosons made of two orbitons 
gives rise to the FQ order and structure transition.
The theoretically predicted multi-orbital-criticality 
presents a unified explanation for 
abovementioned features of iron pnictide superconductors.
\end{abstract}

\address{
$^1$ Department of Physics, Nagoya University and JST, TRIP, 
Furo-cho, Nagoya 464-8602, Japan. 
\\
$^2$ Department of Applied Physics, Nagoya University and JST, TRIP, 
Furo-cho, Nagoya 464-8602, Japan. 
}
 
\pacs{74.70.Xa, 74.20.-z, 74.20.Rp, 74.25.Kc}

\sloppy

\maketitle

\section{Introduction
\label{sec:Intro}}


In iron pnictide superconductors \cite{Hosono},
both spin and orbital degrees of freedom play important roles
on various electronic properties,
such as the high-$T_{\rm c}$ superconductivity, orthorhombic structure
transition, and magnetic transition.
As for the origin of the superconductivity, 
fully-gapped sign-reversing $s$-wave ($s_\pm$-wave) state
had been studied based on spin fluctuation theories
 \cite{Mazin,Kuroki,hirschfeld,chubukov}.
The origin of the spin fluctuations is the intra-orbital nesting 
and the Coulomb interaction.
However, the robustness of $T_{\rm c}$ against randomness in iron pnictides 
indicates the absence of sign-reversal in the superconducting (SC) gap
\cite{Onari-impurity,Sato-imp,Nakajima}.

Later, orbital-fluctuation-mediated $s$-wave state 
without sign reversal ($s_{++}$-wave) had been proposed
based on the Hubbard-Holstein (HH) model 
\cite{Kontani,Saito,Onari}.
The origin of the orbital fluctuations is the inter-orbital nesting 
and the electron-phonon ($e$-ph) interactions due to 
non-$A_{1g}$ optical phonons.
One of the merits of this scenario is the robustness of 
the $s_{++}$-wave state against impurities.
Another merit is that the close relation between $T_{\rm c}$ and the 
crystal structure revealed by Lee \cite{Lee}, 
{\it e.g.}, $T_{\rm c}$ becomes the highest
when the As$_4$ cluster is regular tetrahedron,
is automatically explained \cite{Saito}.
Moreover, orbital-fluctuation-mediated $s_{++}$-wave state scenario is 
consistent with the large SC gap on the $z^2$-orbital band 
in Ba122 systems \cite{Saito}, 
observed by bulk-sensitive laser ARPES measurement \cite{Shimo-Science}.

The ``resonance-like'' hump structure in the neutron 
inelastic scattering \cite{res-exp}
is frequently explained as the spin-resonance due to 
the sign reversal in the SC gap \cite{res-the1,res-the2}.
However, experimental hump structure is well reproduced
in terms of the $s_{++}$-wave SC state, rather than the $s_\pm$-wave SC state,
by taking the suppression 
in the inelastic scattering $\gamma(\w)$ for $|\w|\le3\Delta$
in the SC state (dissipationless mechanism)
\cite{Onari-resonance,Yasui}.
To distinguish between both SC states, measurements of phonon spectral 
function for $|\w|\lesssim 2\Delta$ would be useful 
\cite{Scala}.
In the normal state, prominent
non-Fermi liquid transport phenomena in $\rho$ and $R_{\rm H}$ \cite{Matsuda}
are frequently ascribed to the evidence of spin fluctuations
 \cite{Kontani-review}.
However, they are also explained by the development of 
antiferro-orbital fluctuations \cite{Onari}.

In order to identify the mechanism of superconductivity,
we have to understand the origin of the ordered state,
although it is still unsolved in iron pnictides.
For example, in many heavy fermion superconductors, the SC phase
appears next to the spin-density-wave (SDW) state, indicating the 
occurrence of spin-fluctuation-mediated superconductivity.
In contrast, the ordered state in iron pnictides
is not a simple SDW state: In fact, the tetragonal to orthorhombic 
structure transition occurs at $T_S\sim 100$ K,
usually above the SDW transition temperature $T_{\rm N}$ \cite{review}.
In addition, large imbalance of
$xz$- and $yz$-orbitals at the Fermi level and
reconstruction of the Fermi surfaces (FSs) had been observed
by angle resolved photoemission spectroscopy (ARPES) measurements
\cite{Shimojima,ARPES2,ARPES3}.
These results indicate that the ferro orbital-density-wave (ODW)
is the origin of the orthorhombic structure transition.
The SDW state may originate from the in-plane anisotropy 
in the exchange interaction ($J_{1a}\ne J_{1b}$) associated with the ODW order
\cite{frustration}.

In addition, prominent symmetry breaking $C_4\rightarrow C_2$ is realized 
in detwinned 122 systems even above $T_S$ and $T_{\rm N}$, 
under very small uniaxial pressure.
For example, it is recognized as the large
in-plane anisotropy in the resistivity
 \cite{rho-nematic,Prosorov}
and the optical conductivity \cite{optical} at $T^*\sim200$K, 
which is much higher than $T_S$ and $T_{\rm N}$.
Moreover, the reconstruction of the FSs starts at $T^*$
in detwinned Ba(Fe$_{1-x}$Co$_x$)$_2$As$_2$ \cite{ARPES3}.
The discovery of these ``electronic nematic phase'' 
would indicate that the the ODW fluctuations deveolop
divergently above $T_S$, and the structure transition 
is (almost) the second-order.


Recently, prominent softening of shear modulus
in undoped and under-doped Ba(Fe$_{1-x}$Co$_x$)$_2$As$_2$ had been reported 
by acoustic measurements \cite{Fernandes,Yoshizawa}.
Yoshizawa {\it et al.} \cite{Yoshizawa}
observed all shear moduli $C_{44}$, $C_{66}$ and $C_E$.
Recently, they had also performed systematic measurement for 
$x=0\sim0.225$, and found that only $C_{66}$ shows the prominent softening 
in both under- and over-doped systems \cite{comment4}.
Similar results were reported by Goto {\it et al.} independently \cite{Goto}.
This observation can be explained by the development of
ferro-orbital fluctuations \cite{Saito}
or spin-nematic fluctuations \cite{Fernandes}.
Considering large quadrupole-strain coupling in iron pnictides,
all the observations mentioned above 
suggest the importance of orbital physics, and pose a serious 
challenge for theories of iron pnictide superconductors.
In fact, Goto {\it et al.} have shown that $C_{66}$ is almost 
independent of the magnetic field up to $\sim50$T,
indicating the non-magnetic origin of the softening \cite{Goto}.



Thus, the main features of the iron-pnictide superconductors would be 
summarized  as (i) the orthorhombic (or nematic) transition 
accompanied by remarkable $C_{66}$ softening, 
(ii) emergence of high-$T_{\rm c}$ superconductivity
next to the orthorhombic phase, and (iii) the stripe-type
magnetic order induced by the orthorhombicity.
The unified explanation has not been achieved as far as we know.

In this paper, we develop the orbital fluctuation theory
to explain abovementioned features (i)-(iii)
based on the random-phase-approximation (RPA) and beyond RPA.
In the RPA, large $O_{xz}$-antiferro-quadrupole (AFQ) fluctuations
are produced by the $e$-ph interactions, 
while they do not produce the softening of $C_{66}$ nor $C_{44}$.
If we go beyond the RPA, however,
we find that the $O_{x^2-y^2}$-ferro-quadrupole (FQ) 
fluctuations are brought by the ``two-orbiton process'' 
near the AFQ quantum-critical-point (QCP).
The induced FQ fluctuations cause the softening of $C_{66}$,
and the commensurate ferro-orbital order is realized at $T=T_S$.
It is predicted that the AFQ-QCP is located
at the FQ-QCP, which is the endpoint of the orthorhombic phase.
Near this multi-orbital QCPs,
the superconductivity is mainly caused by the AFQ fluctuations,
and the orthorhombic transition is brought by the FQ fluctuations.
Moreover, the stripe-type antiferro-magnetic state is induced
in the orbital-ordered state,
since the orbital polarization gives strong
in-plane anisotropy in the spin-nesting.
The present study gives a microscopic justification 
for the anisotropic Heisenberg model description
in the SDW state  \cite{frustration,Dai}.


There is a long history in the study of superconductivity
due to charge or orbital fluctuations in multiorbital systems,
starting from the exciton-assisted superconductivity
\cite{Little,Ginzburg,Hirsch,Sawa,Tesa}.
In multiorbital systems, on-site Coulomb interaction
is composed of the intra-orbital term $U$,
inter-orbital one $U'$, and Hund's or exchange term $J$.
Since $U$ is usually larger than $U'\approx U-2J$,
spin fluctuations induced by $U$ give non $s$-wave SC states.
However, Takimoto {\it et al} \cite{Takimoto} had shown that
charge (or orbital) fluctuations induced by $U'$ give 
conventional $s$-wave SC state
if the relations $U'>U$ and $J\sim0$ are assumed.
Although this idea was applied to pnictides \cite{Ohno1},
the relation $U'\sim0.6U$ is realized in many compounds \cite{Miyake}.
Even if $U'\sim0.6U$ and $J\sim0.2U$, 
the present authors had shown that the
orbital-fluctuation-mediated superconductivity is realized by 
quadrupole-quadrupole interaction mediated by non-$A_{1g}$ phonons, 
which works as the ``negative effective exchange $J_{\rm eff}$ for the 
charge sector.'' \cite{Kontani}.
In iron pnictides, orbital fluctuations develop even when the dimensionless 
$e$-ph coupling $\lambda$ due to Fe-ion optical phonons is just $\sim 0.2$, 
according to the RPA \cite{Saito} and FLEX approximation \cite{Onari}.
As for As-ion $A_{1g}$ mode \cite{Ohno1},
orbital fluctuations develop only when $\lambda_{A_{1g}}\sim1$,
which is unrealistic in iron pnictides.

Recently, Yanagi {\it et al}. \cite{Ohno2} 
added the ``orthorhombic phonon''
to the HH model proposed by the present authors \cite{Kontani,Saito,Onari},
and studied the ferro-orbital fluctuations.
However, neither high-$T_{\rm c}$ nor structure transition 
are explained by their theory:
First, the ``orthorhombic-phonon'' is acoustic ($\w_\q\propto|\q|$)
although it is treated as optical in Ref. \onlinecite{Ohno2} incorrectly.
Their theory belongs to the cooperative Jahn-Teller structure transition
due to acoustic phonons like manganites; see details in 
Appendix \ref{sec:ApB}.
In this case, the energy-scale of ``ferro-orbital fluctuations'' 
is too low ($\sim\w_\q$) to explain high-$T_{\rm c}$, since experimental 
$T_{\rm c}$ is much higher than $\w_\q\sim10$K for $|\q|\sim 0.1\pi$ 
\cite{comment,Schrieffer}.
Small orthorhombicity $(a-b)/(a+b)\sim0.003$ in iron pnictides
is also inconsistent with the cooperative Jahn-Teller scenerio.
Second, the derived orbital order is ``incommensurate'' \cite{comment5},
which is inconsistent with the orthorhombic structure transition
and the $C_{66}$ softening.

These problems are resolved in the present theory since
both high-energy AFQ and low-energy ``commensurate'' FQ fluctuations
develop at the same time, without the necessity of fine-tuning
model parameters.
The former (latter) fluctuations give the 
superconductivity (orthorhombic transition).



\section{Model Hamiltonian
\label{sec:model}}
\begin{figure}[!htb]
\includegraphics[width=0.8\linewidth]{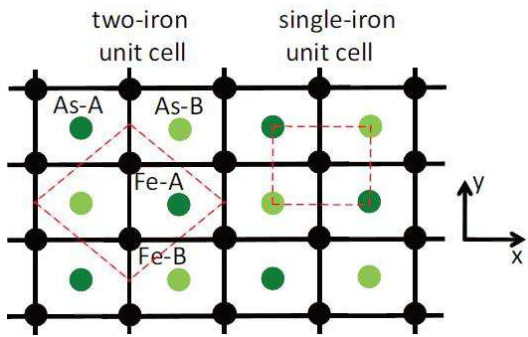}
\caption{
(Color online) 
Crystal structure of FeAs-layer, in which the unit cell is 
given by the two-iron unit cell composed of Fe-A, Fe-B, As-A, and As-B.
The single-iron unit cell is realized by 
applying the ``unfold-gauge transformation''.
}
\label{fig:lattice}
\end{figure}

First, we shortly explain the relation between five-orbital and 
ten-orbital models for iron pnictides.
In this study, we set $x$ and $y$ axes parallel to the nearest Fe-Fe bonds,
and represent the $z^2$, $xz$, $yz$, $xy$, and $x^2-y^2$ $d$-orbitals
in the $xyz$-coordinate as 1,2,3,4, and 5, respectively.
The FSs are mainly composed of $t_{2g}$ orbitals ($l=2\sim4$),
although $e_g$ orbitals also play non-negligible roles
in producing orbital fluctuations \cite{Saito}.
Figure \ref{fig:lattice} shows the crystal structure of FeAs-plane.
Since the As-A (As-B) ions form the upper- (lower-) plane,
the unit cell contains Fe-A and Fe-B, 
which we call the two-iron unit cell.
Since each Fe-ion contains five-orbitals,
the original tight-binding model for iron-pnictides
is given as the ``ten-orbital model''.

In Refs. \onlinecite{Kuroki,Miyake}, 
the authors introduced the gauge transformation
$|2,3\rangle \rightarrow -|2,3\rangle$ only for Fe-B sites,
which we call the ``unfold-gauge transformation''.
Due to this gauge transformation, the unit cell of the kinetic term
is halved to become the single-iron unit cell, 
shown in Figure \ref{fig:lattice}.
In the obtained ``five-orbital model'',
the kinetic term is given as
\begin{eqnarray}
H_{0}=\sum_{ij;lm;\s}t_{lm}^{ij}c_{i, l\s}^\dagger c_{j, m\s} ,
\end{eqnarray}
where $i,j$ denotes the unit cell, 
$l,m=1\sim5$ represent the $d$-orbital, and $\s=\pm1$ is the spin index.
$c_{i, l\s}^\dagger$ is the creation operator of the $d$-electron, and
$t_{lm}^{ij}$ with $i\ne j$ ($i=j$) is the hopping integral (local potential).
This five-orbital model is convenient to study the Eliashberg gap equation
\cite{Kuroki,Kontani}.
In studying the orbital physics, however, 
we have to keep the fact in mind that 
the sign of quadrupole operators ${\hat O}_{xz}$ and ${\hat O}_{yz}$ 
at Fe-B sites are reversed by the unfold-gauge transformation.
By taking care of this fact, we study the softening of shear moduli
based on the five-orbital model hereafter.
In Appendix \ref{sec:10}, we calculate the orbital fluctuations
using the original ten-orbital model,
and make comparison between results of two models.


\begin{figure}[!htb]
\includegraphics[width=0.8\linewidth]{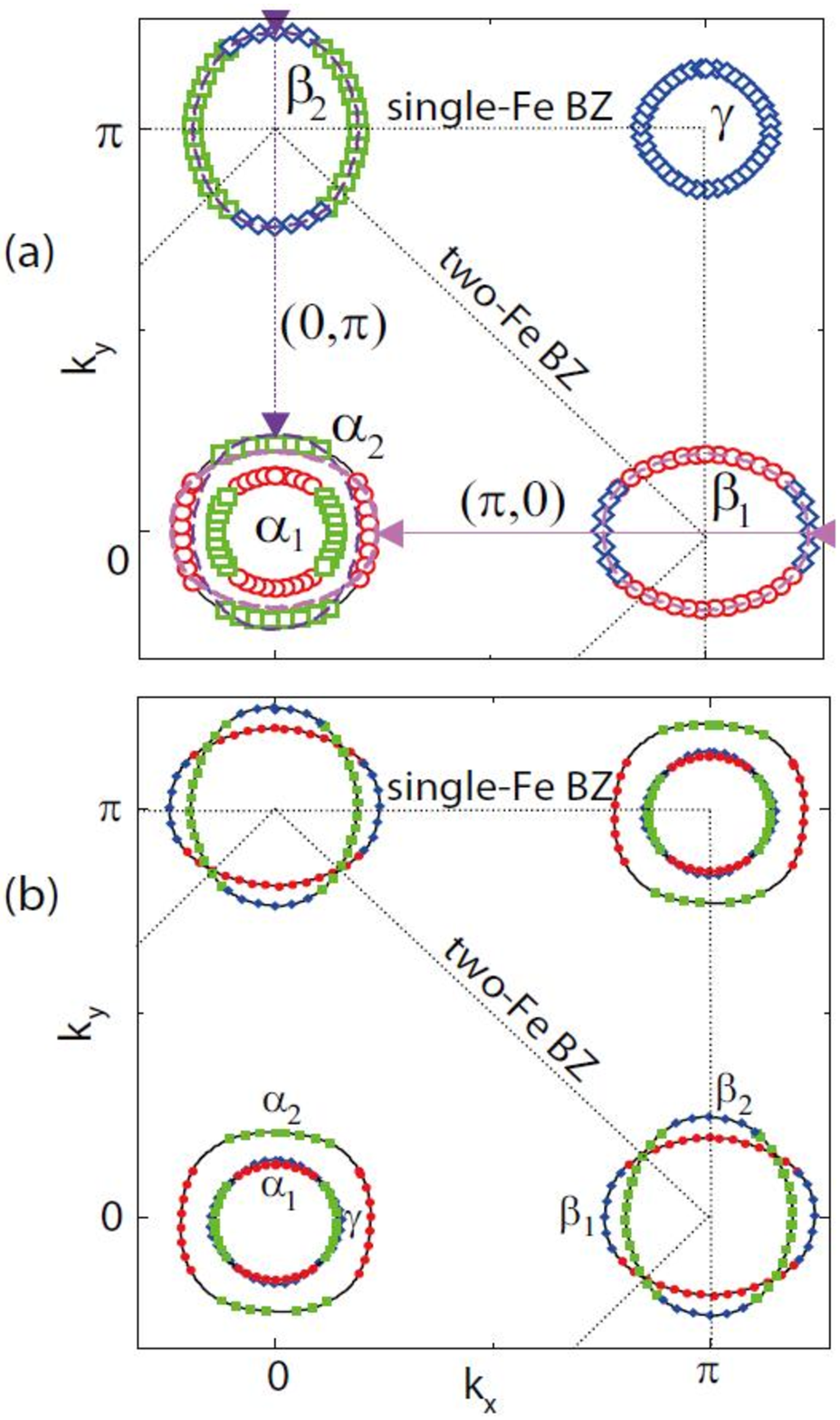}
\caption{
(Color online) 
(a) FSs for $n=6.0$ in the unfolded model.
The colors correspond to $2$ (green), $3$ (red), and $4$ (blue), respectively.
The inter-orbital nesting between FS $\a_2$ (green) and FS $\b_2$ (blue)
causes the AFQ fluctuations.
(b) FSs for the original ten-orbital model.
}
\label{fig:fig1}
\end{figure}

Figure \ref{fig:fig1} shows the FSs in the (a) five-orbital model
and the (b) ten-orbital model.
The FSs in (a) coincide with the FSs in (b)
if we fold the former FSs into the two-iron Brillouin zone (BZ).
We use the hopping parameters for LaOFeAs given in Ref. \onlinecite{Kuroki}.
The colors correspond to $2$ (green), $3$ (red), and $4$ (blue), respectively.
The inter-orbital nesting between the orbital 2 on FS $\a_2$ 
and orbital 4 on FS $\b_2$
causes the most divergent AFQ fluctuations.

Next, we introduce the $e$-ph interaction due to Fe-ion Einstein optical modes.
The Hamiltonian given in Eq. (4) of Refs. \onlinecite{Saito}
is simply rewritten as the following bilinear form
in the $xyz$-coordinate:
\begin{eqnarray}
H_{e-\rm{ph}} =\eta \sum_i\left( {\hat O}^i_{yz}u_x^i +{\hat O}^i_{xz}u_y^i
+{\hat O}^i_{xy}u_z^i \right),
 \label{eqn:e-ph}
\end{eqnarray}
where $\eta=60e^2 a_d^2/7\sqrt{3}R_{\rm Fe-As}^4$;
$a_d$ is the radius of $d$-orbital 
({\it e.g.}, Shannon crystal radius of Fe$^{2+}$ is $0.77$ 
${\buildrel _{\circ} \over {\mathrm{A}}}$), and
$R_{\rm Fe-As}\approx2.4$ ${\buildrel _{\circ} \over {\mathrm{A}}}$.
${\bm u}^i$ is the displacement vector of the $i$-th Fe-ion, and
${\hat O}^i_{\Gamma}$ ($\Gamma=xz,yz,xy$) is the charge quadrupole 
operator given as
\begin{eqnarray}
{\hat O}^i_{\Gamma}\equiv \sum_{lm}^\pm o^{l,m}_\Gamma {\hat m}_{l,m}^i ,
\end{eqnarray}
where ${\hat m}_{l,m}^i \equiv \sum_\s c_{i,l\s}^\dagger c_{i,m\s}$,
and the coefficient is defined as
$o^{l,m}_{xz}= 7\langle l|{\hat x}{\hat z}|m \rangle$
for $\Gamma=xz$, where ${\hat \x}=x/r$ and so on.
The non-zero coefficients are given as
\begin{eqnarray}
&& o^{2,5}_{xz}=o^{3,4}_{xz}=\sqrt{3}o^{1,2}_{xz}=1,
 \label{eqn:oxz}\\
&& -o^{3,5}_{yz}=o^{2,4}_{yz}=\sqrt{3}o^{1,3}_{yz}=1,
  \label{eqn:oyz}\\
&& o^{2,3}_{xy}=-\sqrt{3}o^{1,4}_{xy}/2=1.
 \label{eqn:oxy}
\end{eqnarray}
%
Be careful not to confuse ${\hat O}_{xz}$ with the $xz$-orbital operator.
Other two quadrupole operators are $O_{z^2}$ and $O_{x^2-y^2}$,
whose coefficients are respectively defined as
$o^{l,m}_{x^2-y^2}= (7/2)\langle l|({\hat x}^2-{\hat y}^2)|m \rangle$
and $o^{l,m}_{z^2}= (7/2\sqrt{3})\langle l|(3{\hat z}^2-1)|m \rangle$.
(They are written as $O_2^0$ and $O_2^2$ in literatures.)
The non-zero coefficients are given as
\begin{eqnarray}
&& o^{2,2}_{x^2-y^2}=-o^{3,3}_{x^2-y^2}=-(\sqrt{3}/2)o^{1,5}_{x^2-y^2}=1,
 \label{eqn:ox2y2}\\
&& o^{1,1}_{z^2}=2o^{2,2}_{z^2}=2o^{3,3}_{z^2}=-o^{4,4}_{z^2}=-o^{5,5}_{z^2}=2/\sqrt{3}.
 \label{eqn:oz2} 
\end{eqnarray}
Expect for $\Gamma=z^2$, all the matrix elements of ${\hat o}_\Gamma$ 
with respect to the $t_{2g}$-orbital ($2\sim4$) are $\pm 1$.

Here, we derived the $e$-ph interaction based on the point-charge model.
Although $e$-ph interaction is also induced by the change 
in the $d$-$p$ hopping, as discussed in Ref. \onlinecite{Yada},
we expect it is small since the weight of $p$-electron
on the Fermi surface is just $\sim5$\% in iron pnictides.
Fortunately, because of the Wigner-Eckart theorem, 
the matrix elements of the local 
quadrupole-phonon interaction is always given by 
the quadrupole operator ${\hat O}^i_{\Gamma}$,
independently of the details of the interaction.
Since the magnitude of the hexadecapole-phonon interaction is 
$(a_d/R_{\rm Fe-Fe})^2\sim0.1$ times that of the quadrupole-phonon interaction,
we can safely use Eq. (\ref{eqn:e-ph}).

\begin{figure}[!htb]
\includegraphics[width=0.8\linewidth]{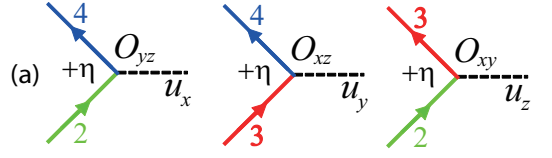}
\caption{
(Color online) 
Inter-orbital scattering processes due to the 
$e$-ph interaction by $u_\mu$ ($\mu=x,y,z$)
within the $t_{2g}$-orbitals ($l=2\sim4$).
}
\label{fig:e-ph}
\end{figure}

Equation (\ref{eqn:e-ph}) means that the displacement $u_x$ 
produces the quadrupole potential ${\hat O}_{yz}$,
which causes the scattering of electrons between orbitals 2 and 4.
The $e$-ph interactions in Eq. (\ref{eqn:e-ph}) within $t_{2g}$-orbitals
are shown in Fig. \ref{fig:e-ph}.
Then, the phonon-mediated el-el interaction $V_{\rm{el-el}}^{\rm ph}$ 
is obtained by taking the contraction of ${\bm u}^i$, which gives
the local phonon Green function 
$D(\tau)\equiv \langle T_{\tau} u_{\mu}^i ( \tau ) u_{\mu}^i ( 0 ) \rangle $ 
$( \mu = x,y,z )$.
By taking the Fourier transformation, we obtain
\begin{equation}
D(\omega_l) = \frac{2 \langle u^2 \rangle_0 \omega_D}{\omega_l^2 + \omega_D^2},
\label{eqn:D}
\end{equation}
where $\w_l=2\pi Tl$ is the Boson Matsubara frequency, 
$\omega_D$ is the optical phonon frequency, and 
$\sqrt{\langle u^2 \rangle_0} = \sqrt{ 1 / 2M_{\text{Fe}} \omega_D}$ is the 
uncertainty in position for Fe ions; $\sqrt{\langle u^2 \rangle_0} = 0.044$ 
${\buildrel _{\circ} \over {\mathrm{A}}}$ for $\w_{\rm D}=0.02$ eV \cite{Saito}.
Then, $V_{\rm{el-el}}^{\rm ph}$ is expressed as the 
following quadrupole-quadrupole interaction:
%
\begin{eqnarray}
V_{\rm{el-el}}^{\rm ph}=-g(\w_l) \sum_i\left\{ 
{\hat O}^i_{yz} \cdot{\hat O}^i_{yz} + {\hat O}^i_{xz} \cdot{\hat O}^i_{xz}
+ {\hat O}^i_{xy} \cdot{\hat O}^i_{xy} \right\},
 \nonumber \\
 \label{eqn:Hint}
\end{eqnarray}
where $g(\w_l)=g \cdot \w_{\rm D}^2/(\w_l^2+\w_{\rm D}^2)$, and 
$g$ is the phonon-mediated el-el interaction at zero frequency; 
$g=0.34$eV in the present point-charge model \cite{Kontani,Saito}.

\section{Random-phase-approximation
\label{sec:RPA}}

Now, we explain the RPA for the five-orbital HH model \cite{Takimoto}.
The irreducible susceptibility in the five orbital model is given by
\begin{equation} 
\chi^0_{ll',mm'} \left( q \right) = - \frac{T}{N} \sum_\k G_{lm}^0 
\left( k+ q \right) G_{m' l'}^0 \left( k \right),
\end{equation}
where $\hat{G}^0 ( k ) = [ i \epsilon_n + \mu - \hat{H}^0_{\bm{k}} ]^{-1}$
is the \textit{d} electron Green function in the orbital basis,
$q = ( \bm{q}, \omega_l )$, $k=( \bm{k} , \epsilon_n)$,
and $\epsilon_n = (2n + 1) \pi T$ is the fermion Matsubara frequency.
$\mu$ is the chemical potential, and $\hat{H}^0_{\bm{k}}$ is the kinetic term.
Then, the susceptibilities for spin and charge sectors in the RPA are 
given as \cite{Takimoto}
\begin{gather} 
\hat{\chi}^{s} \left( q \right) = \frac{\hat{\chi}^0 \left( q \right)}{1 - \hat{\Gamma}^{s} \hat{\chi}^0 \left( q \right)}, 
\label{eqn:chis}\\
\hat{\chi}^{c} \left( q \right) = \frac{\hat{\chi}^0 \left( q \right)}{1 - \hat{\Gamma}^{c} (\omega_l) \hat{\chi}^0 \left( q \right)},
\end{gather}
where the bare four-point vertices ${\hat \Gamma}^{s,c}$ are
\begin{eqnarray}
\Gamma_{l_{1}l_{2},l_{3}l_{4}}^s = \begin{cases}
U, & l_1=l_2=l_3=l_4 \\
U' , & l_1=l_3 \neq l_2=l_4 \\
J, & l_1=l_2 \neq l_3=l_4 \\
J' , & l_1=l_4 \neq l_2=l_3
\end{cases}
\label{eqn:Gamma-s}
\end{eqnarray}
\begin{equation}
\hat{\Gamma}^c ( \omega_l )= -\hat{C} - 2\hat{V}_{\rm el-el}^{\rm ph}( \w_l ),
\label{eqn:Gc}
\end{equation}
\begin{eqnarray}
C_{l_{1}l_{2},l_{3}l_{4}} = \begin{cases}
U, & l_1=l_2=l_3=l_4 \\
-U'+2J , & l_1=l_3 \neq l_2=l_4 \\
2U' - J, & l_1=l_2 \neq l_3=l_4 \\
J' , & l_1=l_4 \neq l_2=l_3
\end{cases}
\label{eqn:Gamma-c}
\end{eqnarray}
%
In Eq. (\ref{eqn:Gc}), $(V_{\rm el-el}^{\rm ph})_{l_1,l_2,l_3,l_4}
=-g(\w_l)\sum_\Gamma^{xz,yz,yx}o^{l_1,l_2}_\Gamma o^{l_3,l_4}_\Gamma$.
Here, we neglect the ladder-diagram for phonon-mediated interaction
because of the relation $\omega_D \ll W_{\mathrm{band}}$ \cite{Kontani}.

In the RPA,
the enhancement of the spin susceptibility $\hat{\chi}^{s}$ 
is mainly caused by the intra-orbital Coulomb interaction $U$,
using the ``intra-orbital nesting''.
On the other hand,
the enhancement of $\hat{\chi}^{c}$ in the present model is caused by 
the phonon-induced quadrupole-quadrupole interaction in Eq. (\ref{eqn:Hint}),
utilizing the ``inter-orbital nesting'' in the present model.
The SDW (ODW) state is realized when the spin (charge) Stoner factor $\a_{s(c)}$,
which is the maximum eigenvalue of $\hat{\Gamma}^{s(c)} \hat{\chi}^0(\bm{q},0)$,
is unity.
When $n=6.05$,
the critical value of $U$ is $U_c=1.26$ eV, and 
the critical value of $g$ (at $U=0$) is $g_c=0.233$ eV. 
Smallness of $g_{\rm c}$ in iron pnictides originates from
the better inter-orbital nesting.
Hereafter, we set the unit of energy as eV unless otherwise noted.

\begin{figure}[!htb]
\includegraphics[width=0.8\linewidth]{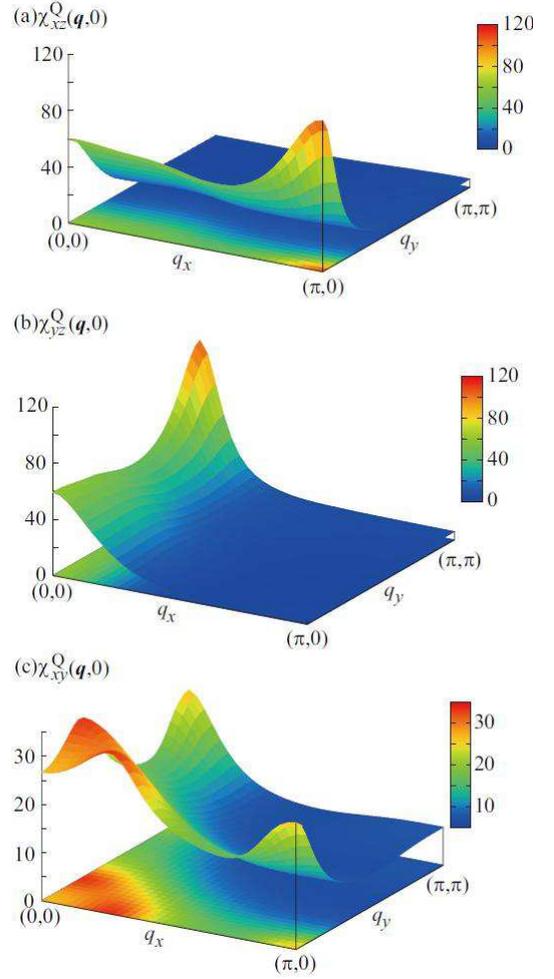}
\caption{
(Color online) 
Quadrupole susceptibilities in the five-orbital model for
(a) $\chi_{xz}^{Q}(\q)$, (b) $\chi_{yz}^{Q}(\q)$, and 
(c) $\chi_{xy}^{Q}(\q)$, respectively.
The used model parameters are $n=6.05$, $T=0.05$, and $g=0.22$
($\a_c=0.98$).
The correlation length in (a) or (b) is derived as
$\xi=\pi/\Delta q\sim3$, where $\Delta q$ is the half-width of the peak.
Therefore, we obtain the relations $6\xi^2\sim (1-\a_c)^{-1}$
and $c\xi^2\sim 2(1-\a_c)^{-1}\sim 12\xi^2$.
}
\label{fig:chic}
\end{figure}

Here, we introduce the diagonal charge quadrupole 
susceptibilities in the five-orbital model as
\begin{eqnarray}
\chi_{\Gamma}^{Q}(q)&=&
\sum_{ll'}\sum_{mm'}o_\Gamma^{ll'}\chi_{ll'mm'}^c(q)o_\Gamma^{mm'},
 \label{eqn:chiQ} 
\end{eqnarray}
for $\Gamma=xz,yz,xy$.
Their momentum dependence at zero frequency
is shown in Fig. \ref{fig:chic} for $\a_c=0.98$.
More generally, the quadrupole susceptibility is defined as
\begin{eqnarray}
\chi_{\Gamma,\Gamma'}^{Q}(q)=
\sum_{ll'}\sum_{mm'}o_\Gamma^{ll'}\chi_{ll'mm'}^c(q)o_{\Gamma'}^{mm'} .
\label{eqn:offdiagonal}
\end{eqnarray}
%
However, its off-diagonal terms with $\Gamma\ne\Gamma'$ 
are negligibly small in the present model in the ``$xyz$-coordinate''.
In this approximation, in the absence of Coulomb interaction,
the quadrupole susceptibility in the RPA is given as
\begin{eqnarray}
\chi_{\Gamma}^{Q}(q)\approx 
 \chi_{\Gamma}^{Q,0}(q)/(1-2g\chi_{\Gamma}^{Q,0}(q))
 \label{eqn:chiGU0} ,
\end{eqnarray}
where $\chi_{\Gamma}^{Q,0}$ is the irreducible quadrupole susceptibility.
Considering the fact that $\chi_{ll',mm'}^0(q)$ takes large value for 
$l=m$ and $l'=m'$, we obtain $\chi_{xz}^{Q,0}(q)\approx 
2\chi_{25,25}^0(q)+2\chi_{34,34}^0(q)+(2/3)\chi_{12,12}^0(q)$.
Since $\chi_{xz}^{Q,0}(q)\approx 2.5$ at $\q=(\pi,0)$,
the critical value of $g$ is $g_c\sim0.2$ in the present model.
We stress that the relations
$\chi_{\Gamma,\Gamma'}^{Q}(q)=\chi_{\Gamma}^{Q}(q)\delta_{\Gamma,\Gamma'}$
and Eq. (\ref{eqn:chiGU0}) holds exactly for $q=({\bm 0},\w_l)$.
We utilize this relation in calculating the shear modulus.

As for the contributions by $t_{2g}$-orbitals ($l=2\sim4$), 
$\chi_{xz}^{Q}(q)\propto \chi_{34,34}^c(q)$,
$\chi_{yz}^{Q}(q)\propto \chi_{24,24}^c(q)$, and
$\chi_{xy}^{Q}(q)\propto \chi_{23,23}^c(q)$.
In Fig. \ref{fig:chic} (a), $\chi_{xz}^{Q}(\q)$ has the highest peak 
at $\q=(\pi,0)$, which is given by the  
inter-orbital nesting between orbital 3 on FS $\a_2$
and orbital 4 on FS $\b_1$ in the five-orbital model
in Fig. \ref{fig:fig1} (a).
Also, $\chi_{yz}^Q(\q)$ has the highest peak at $\q=(0,\pi)$
in Fig. \ref{fig:chic} (b),
due to the inter-orbital nesting between orbital 2 on FS $\a_2$
and the orbital 4 on FS $\b_2$.
We will see that $\chi_{xz(yz)}^Q(\q)$ is modified by the unfolding procedure.

Also, $\chi_{xy}^Q(\q)$ in Fig. \ref{fig:chic} (c)
is given by the inter-orbital nesting between orbital 2 and 3, 
due to the out-of-plane oscillations of Fe-ions.
The inter-band and intra-band scattering processes produce the
enhancement of $\chi_{xy}^Q(\q)$ at $\q=(\pi,0), (0,\pi)$
and $\q={\bm 0}$, respectively.
We note that $\chi_{xy}^Q(\q)$ is not affected by the unfolding procedure.

\begin{figure}[!htb]
\includegraphics[width=0.8\linewidth]{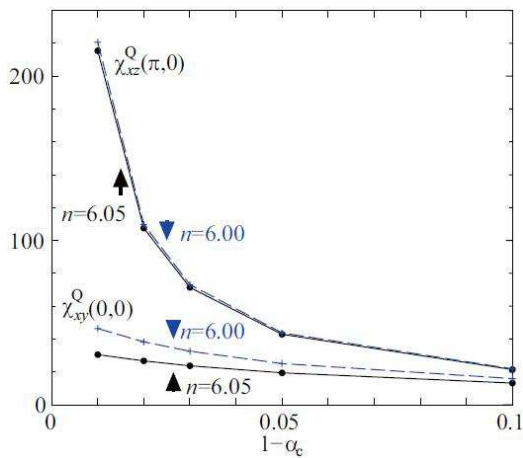}
\caption{
(Color online) 
Quadrupole susceptibilities as a function of $\a_c$ 
for $n=6.0$ and $n=6.05$ given by RPA.
Model parameters are $U=0.8$ and $T=0.05$.
We can recognize the relation 
$\chi_{xz}^Q({\bm Q})\propto (1-\a_c)^{-1} \propto (g_{\rm c}-g)^{-1}$,
which diverges at $\a_c=1$ or $g=g_{\rm c}$.
}
\label{fig:chic-ac}
\end{figure}

Figure \ref{fig:chic-ac} shows both
$\chi_{xz}^Q(\q)$ at $\q=(\pi,0)$ and $\chi_{xy}^Q({\bm 0})$
as a function of $\a_c$ for $n=6.0$ and $n=6.05$ given by the RPA.
We see that $\chi_{xz}^Q(\q)$ develops divergently 
in proportion to $(1-\a_c)^{-1}\propto (g_c-g)^{-1}$,
while $\chi_{xy}^Q({\bm 0})$ shows an enhanced but saturated value
even at $g=g_{\rm c}$.

\begin{figure}[!htb]
\includegraphics[width=0.88\linewidth]{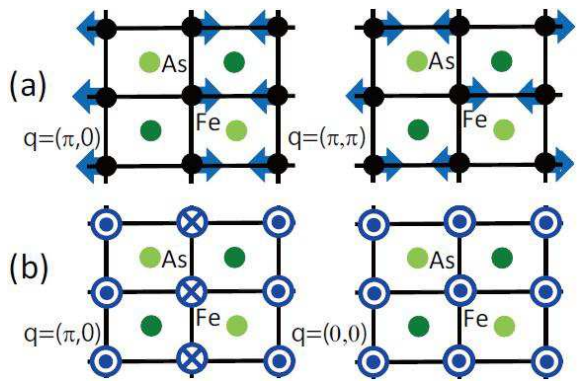}
\caption{
(Color online) 
(a) Fe-ion in-plane optical phonons 
with momentum $\q=(\pi,0)$ and $\q=(\pi,\pi)$.
(b) Fe-ion out-of-plane optical phonons 
with momentum $\q=(\pi,0)$ and $\q=(0,0)$.
}
\label{fig:optical}
\end{figure}

In Appendix \ref{sec:10},
we will calculate $\chi_{\Gamma}^{Q}(q)$ in the ten-orbital model,
and make comparison to Fig. \ref{fig:chic} in the five-orbital model:
Although both results coincide for $\Gamma=xy,z^2,x^2-y^2$,
they are different for $\Gamma=xz,yz$.
The reason is that the signs of $O_{xz/yz}$ at Fe-B sites are changed 
by applying the ``unfold-gauge transformation''.
As we will explain in Appendix \ref{sec:10}, the development of 
$\chi_{xz/yz}^{Q}({\bm 0},0)$ in Fig. \ref{fig:chic} is the artifact 
of the unfold-gauge transformation.
For this reason,
the correct AFQ susceptibility in the ten-orbital model is given as 
$\chi_{xz/yz}^{Q}(\q)=\chi_{xz/yz}^{Q,\rm{5-orbital}}(\q+(\pi,\pi))$.
The optical modes that give the enhancements of 
$\chi_{yz}^{Q}(\q)$ at $\q=(\pi,0)$ and $(\pi,\pi)$ 
in the ten-orbital model 
($\q=(0,\pi)$ and $(0,0)$ in the five-orbital model)
are caused by the in-plane $u_{x}$
oscillations shown in Fig.\ref{fig:optical} (a).
Also, the enhancements of $\chi_{xy}^{Q}(\q)$ at $\q=(\pi,0)$
and $(0,0)$ are caused by the out-of-plane $u_z$
oscillations in Fig.\ref{fig:optical} (b).

\section{Acoustic phonons
\label{sec:acoustic}}
In previous sections, we studied the $e$-ph interaction due to 
optical phonons, and calculated the quadrupole susceptibilities by the RPA.
To obtain the shear modulus, 
we also need the knowledge on the $e$-ph interaction due to 
acoustic phonons with momentum $\q\approx0$.
In fact, shear modulus is proportional to the square of 
the acoustic phonon velocity, which is renormalized 
by the electron-acoustic phonon interaction in the presence of strong
quadrupole fluctuations.
In this section, we derive the $e$-ph interaction due to 
acoustic phonons with $\q\approx0$.
Hereafter, we use the unit $\hbar=1$, and take 
the nearest-neighbor Fe-Fe distance $a_{\rm Fe-Fe}$ as the unit of length.

\begin{figure}[!htb]
\includegraphics[width=0.9\linewidth]{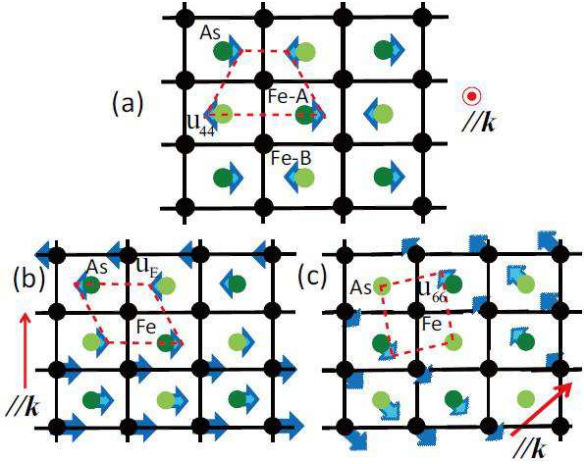}
\caption{
(Color online) 
Displacement vectors ${\bm u}_{\rm As}$ and ${\bm u}_{\rm Fe}$ 
in the transverse acoustic modes
that couple with (a) $C_{44}$, (b) $C_E$, and (c) $C_{66}$.
$C_{66}$ mode corresponds to the orthorhombic structure transition.
}
\label{fig:accoustic}
\end{figure}

Figure \ref{fig:accoustic} shows the transverse acoustic modes that are
related to (a) $C_{44}$, (b) $C_E=(C_{11}-C_{12})/2$, and (c) $C_{66}$
\cite{Yoshizawa}.
Now, we calculate the $e$-ph interaction based on the point-charge model,
by following the procedure in Ref. \onlinecite{Kontani} for the optical phonons.
The quadrupole potential energies at Fe-site caused by the 
transverse acoustic phonons in Fig. \ref{fig:accoustic} are given as
\begin{eqnarray}
V_{44}&=& -\frac{3e^2}{R_{\rm Fe-As}^4}\frac{8}{\sqrt{3}} xz\cdot 
{\tilde u}_{44}
 \label{eqn:V44} ,\\
V_{E}&=& -\frac{3e^2}{R_{\rm Fe-As}^4}\frac{8}{\sqrt{3}} xy\cdot
{\tilde u}_{E}
 \label{eqn:VE} ,\\
V_{66}&=& \frac{3e^2}{R_{\rm Fe-As}^4}\sqrt{6} (x^2-y^2)\cdot 
{\tilde u}_{66}
 \label{eqn:V66} ,
\end{eqnarray}
for both Fe-A and Fe-B sites, where 
$(x,y,z)$ is the coordinates of $d$-electron.
${\tilde {\bm u}}_{\phi}\equiv {\bm u}_{\phi}-{\bm u}_{\rm Fe}$ ($\phi=44,66,E$)
is the relative displacements
of the nearest As ions from the center Fe ion;
$u_{\phi}$ ($u_{\rm Fe}$) is the displacement vector of 
the As- (Fe-) ion we are considering from the original position.
Note that the shear strain tensors are given as
$\e_{44(E)}={\tilde u}_{44(E)}/(a_{\rm Fe-Fe}/2)=2{\tilde u}_{44(E)}$ and 
$\e_{66}={\tilde u}_{66}/(a_{\rm Fe-Fe}/\sqrt{2})=\sqrt{2}{\tilde u}_{66}$.

The corresponding operators in the ten-orbital model
are respectively given as
\begin{eqnarray}
{\hat V}_{44}&=& -\frac{3e^2a_d^2}{R_{\rm Fe-As}^4}\frac{8}{7\sqrt{3}} 
 {\hat O}_{xz}\cdot {\tilde u}_{44}
 \label{eqn:V44O} ,\\
{\hat V}_{E}&=& -\frac{3e^2a_d^2}{R_{\rm Fe-As}^4}\frac{8}{7\sqrt{3}} 
 {\hat O}_{xy}\cdot {\tilde u}_{E}
 \label{eqn:VEO} ,\\
{\hat V}_{66}&=& \frac{3e^2a_d^2}{R_{\rm Fe-As}^4}\frac{2\sqrt{6}}{7} 
 {\hat O}_{x^2-y^2}\cdot {\tilde u}_{66}
 \label{eqn:V66O} .
\end{eqnarray}
Therefore, the acoustic modes in Fig. \ref{fig:accoustic} (a)-(c)
couple with the quadrupole susceptibilities at $\q\approx0$;
$\chi_{xz}^Q(0)$, $\chi_{xy}^Q(0)$, and 
$\chi_{x^2-y^2}^Q(0)$ for $\phi=44$ $E$, and $66$ 
in the ten-orbital model, respectively.

To study the softening in the five-orbital models,
we have to perform the ``unfold-gauge transformation''
for Eqs. (\ref{eqn:V44O})-(\ref{eqn:V66O}).
Under the gauge transformation,
Eqs. (\ref{eqn:VEO}) and (\ref{eqn:V66O}) are invariant,
while Eq. (\ref{eqn:V44O}) is changed to 
\begin{eqnarray}
{\hat V}'_{44}&=& \mp \frac{3e^2a_d^2}{R_{\rm Fe-As}^4}\frac{8}{7\sqrt{3}} 
 {\hat O}_{xz}\cdot {\tilde u}_{44} ,
 \label{eqn:V44O-2}
\end{eqnarray}
where the $-(+)$ sign corresponds to Fe-A (Fe-B) site.
In the ``five-orbital model'', therefore, 
the softening of $C_{E}$ and $C_{66}$ are caused by
$\chi_{xy}^Q({\bm 0},0)$ and $\chi_{x^2-y^2}^Q({\bm 0},0)$, respectively,
while the softening of $C_{44}$ is caused by $\chi_{xz}^Q((\pi,\pi),0)$.
Therefore, the softening in shear modulus ($C_{66}$ and $C_{44}$)
does not occur within the RPA \cite{comment2}.

Next, we derive the effective el-el interaction due to $\k\rightarrow0$ 
transverse acoustic modes.
In the case of $\k=k\cdot(1,1)/\sqrt{2}$ and $k\ll1$
shown in Fig. \ref{fig:accoustic} (c),
the displacement operator for the As site at ${\bm R}_s$ is
\begin{eqnarray}
u_s= \sum_\k \sqrt{\frac{1}{2NM\w_\k}}
\left[ a_\k e^{i\k{\bm R}_s}+a_\k^\dagger e^{-i\k{\bm R}_s} \right] ,
\end{eqnarray}
where 
$a_\k$ and $a_\k^\dagger$ satisfy the commutation relation
$[a_\k,a_{\k'}^\dagger]=\delta_{\k,\k'}$, $M$ is the mass of As-ion, and
$\w_\k=v_\k |\k|$; $v_\k$ is the bare acoustic phonon velocity.
The Fourier transformation of $u_s$ is given as
$u_\k= \sqrt{\frac{1}{2M\w_\k}}(a_\k+a_{-\k}^\dagger)$.
Then, the Fourier transformation of the acoustic-phonon Green function 
$D_\k(\tau)\equiv \langle T_{\tau} u_\k^i ( \tau ) u_\k^i ( 0 ) \rangle$ 
is 
\begin{eqnarray}
D_\k(\w_n)=\frac{2\w_\k}{\w_n^2+\w_\k^2}\langle{u}_\k^2\rangle_0 ,
 \label{eqn:Dq}
\end{eqnarray}
where $\langle {u}_\k^2 \rangle_0 =(1/2M\w_\k)$.

As understood in Fig. \ref{fig:accoustic} (c),
the relative displacement with the origin at the Fe-ion, 
${\tilde u}_s$, is given as $\frac12 (u_{s'}-u_s)$, where
${\bm R}_{s'}=(0.5,0.5)$ and ${\bm R}_{s}=(-0.5,-0.5)$
with the origin at the Fe-ion.
Considering that $({\bm R}_{s'}-{\bm R}_{s})\cdot \k=\sqrt{2}k$,
its Fourier transformation is given as 
\begin{eqnarray}
{\tilde u}_\k\equiv \sum_s {\tilde u}_s e^{-i \k{\bm R}_s}
\sim \frac{i k}{\sqrt{2}}u_\k .
\end{eqnarray}
To calculate the shear modulus,
we need the quadrupole susceptibility in the ``$k$-limit'', 
in which we put $\w=0$ first, and take the limit $k\rightarrow0$ later
\cite{Kontani-VV}.
For this purpose, we derive the effective el-el interaction
due to the transverse acoustic phonon in the $k$-limit.
The el-el interaction due to the phonon in Fig. \ref{fig:accoustic} (c)
is given by the second-order term of Eq. (\ref{eqn:V66O}).
Using the relation $D_\k(0)\k^2=2(1/2M\w_\k)\k^2/\w_\k=1/Mv_\k^2$,
it is given as
\begin{eqnarray}
H_{66}&=&-g_{66}\sum_{\k\k',ll'mm',\s\s'}o_{x^2-y^2}^{ll'}o_{x^2-y^2}^{mm'}
 \nonumber\\
& &\times c_{l\k\s}^\dagger c_{l'\k\s}c_{m\k'\s'}^\dagger c_{m'\k'\s'},
 \label{eqn:H66} \\
g_{66}&=& \frac{B^2}{R_{Fe-As}^2}\frac{1}{Mv_\k^2},
\end{eqnarray}
where $\displaystyle B\equiv \frac{3e^2}{R_{\rm Fe-As}}
\left(\frac{a_d}{R_{\rm Fe-As}}\right)^2\frac{2\sqrt{3}}{7}$;
$B=0.95$ eV for $a_d=0.77$ ${\buildrel _{\circ} \over {\mathrm{A}}}$
(=Shannon crystal radius of Fe$^{2+}$) 
and $R_{\rm Fe-As}=2.4$ ${\buildrel _{\circ} \over {\mathrm{A}}}$.
Therefore,
\begin{eqnarray}
g_{66}= \eta_{66}^2 C_{66,0}^{-1} ,
\end{eqnarray}
where $C_{66,0}\equiv Mv_\k^2$ is the bare shear modulus,
and $\eta_{66}= BR_{\rm Fe-As}^{-1}$ is the quadrupole-strain coupling constant.
If we put $v_\k\sim0.024$ eV  ${\buildrel _{\circ} \over {\mathrm{A}}}$
($v_\k\sim0.018$ eV  ${\buildrel _{\circ} \over {\mathrm{A}}}$)
according to the first principle study \cite{Boeri}, 
we obtain $g_{66}=0.12$eV ($g_{66}=0.21$eV).
On the other hand, we obtain $g=0.34$eV for the Fe-ion
optical phonons with $\w_{\rm D}=0.02$eV \cite{Saito}.
Thus, $g_{66}/g=1/2\sim1/3$ in the present point charge model.


In the same way, we also derive the el-el interactions due to the 
acoustic phonon with $\k=k\cdot(0,1)$, shown in Fig. \ref{fig:accoustic} (b).
For this mode, the relative displacement ${\tilde u}_\k$ is given as
${\tilde u}_s\equiv \frac12 (u_{s'}-u_s)$
with ${\bm R}_{s'}=(0.5,0.5)$ and ${\bm R}_{s}=(0.5,-0.5)$.
Since $({\bm R}_{s'}-{\bm R}_{s})\cdot\k=k$,
its Fourier transformation is given by
\begin{eqnarray}
{\bar u}_\k\equiv \sum_s {\tilde u}_s e^{-i \k{\bm R}_s}
\sim \frac{i k}{2} u_\k .
\end{eqnarray}
Then, the phonon-mediated el-el interactions
are given by the second-order terms of Eq. (\ref{eqn:VEO}).
As a result,
the el-el interactions due to phonons in Fig. \ref{fig:accoustic} (b) 
is given as
\begin{eqnarray}
H_{E}&=&-g_{E}\sum_{\k\k',ll'mm',\s\s'}o_{xy}^{ll'}o_{xy}^{mm'}
 \nonumber\\
& &\times c_{l\k\s}^\dagger c_{l'\k\s}c_{m\k'\s'}^\dagger c_{m'\k'\s'},
\end{eqnarray}
where 
$\displaystyle g_{E}=\frac{B'^2}{R_{Fe-As}^2}\frac{1}{Mv_\k^2}$,
and $\displaystyle B'\equiv \frac{3e^2}{R_{\rm Fe-As}}
\left(\frac{a_d}{R_{\rm Fe-As}}\right)^2\frac{4}{7\sqrt{3}}$.
In the same way, we obtain $g_{44}=g_{E}$.
Therefore, $g_{\phi}=\eta_{\phi}^2 C_{\phi,0}^{-1}$ 
and $\eta_{\phi}^2=0.44\eta_{66}^2$ for $\phi=44, E$.
In conclusion, $g_{E}=g_{44}=0.44g_{66}$ 
if $v_\k$ is equivalent for all modes.

\section{Softening of shear moduli
\label{sec:shear}}

\subsection{Softening due to one-orbiton process; the RPA}

\begin{figure}[!htb]
\includegraphics[width=0.9\linewidth]{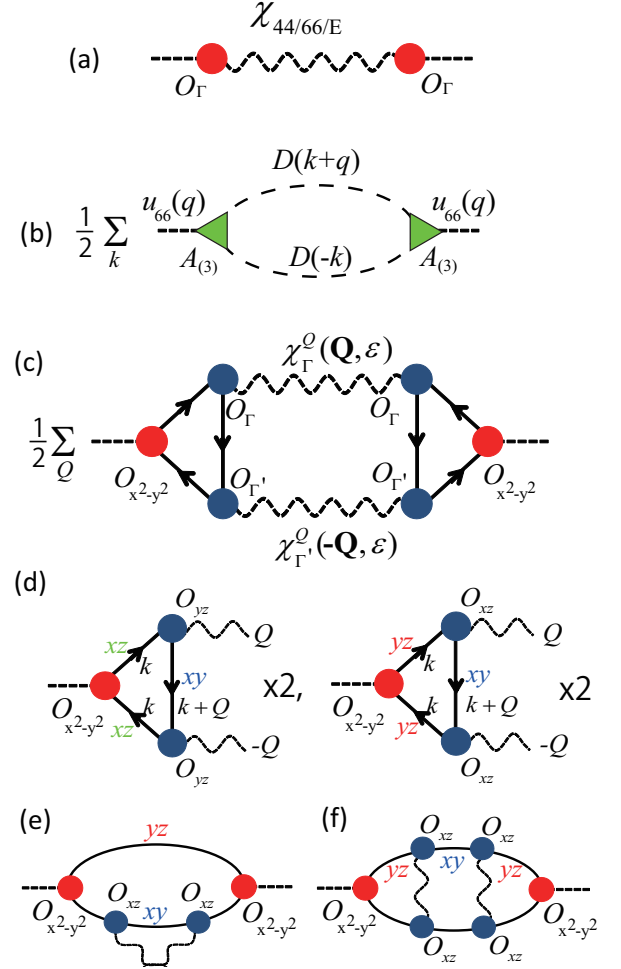}
\caption{
(Color online) 
(a) Diagrammatic expression for the shear modulus susceptibilities
$\chi_{\phi}$ ($\phi=44,66,E$) in the RPA.
(b) The second-order term with respect to the third-order anharmonic
phonon-phonon interaction $A_{(3)}u_x^2u_{66}$.
This diagram represents the virtual process in which
an acoustic phonon with $\q=0$ breaks into 
two optical phonons conserving the total momentum.
(c) Two-orbiton term for the shear modulus susceptibilities
$\chi_{66}^{\rm TO}$, which is the irreducible susceptibility of
$\chi_{x^2-y^2}({\bm 0},0)$.
This term gives the softening of $C_{66}$.
(d) Dominant contribution for the three-point vertex 
$A_{x^2-y^2}(\Gamma,\Gamma;\q)$ for $\Gamma=xz$ and $yz$.
(e) Self-energy correction due to $\chi^Q_{xz}(q)$.
(f) Second-order Maki-Thompson-type vertex correction 
with respect to $\chi^Q_{xz}(q)$.
}
\label{fig:AL}
\end{figure}

Here, we calculate the shear modulus given by the one-orbiton process
using the RPA.
For this purpose, we introduce the following shear modulus susceptibilities
in the five-orbital model,
in the absence of $e$-ph interaction due to $\q\approx0$ acoustic phonon:
\begin{eqnarray}
\chi_E&=& 2\chi_{xy}^{Q}({\bm 0},0), 
 \label{eqn:chiE-def} \\
\chi_{44}&=& 2\chi_{xz}^{Q}((\pi,\pi),0), 
  \label{eqn:chi44-def} \\
\chi_{66}&=& 2\chi_{x^2-y^2}^{Q}({\bm 0},0),
  \label{eqn:chi66-def} 
\end{eqnarray}
where the factor 2 comes from the spin degeneracy.
They are schematically depicted in Fig. \ref{fig:AL} (a).
Note that $\chi_{44}= 2\chi_{xz}^{Q}({\bm 0},0)$ in the ten-orbital model.
According to Sec. 2 in Ref. \onlinecite{Thal},
the shear modulus is given by the second derivative 
of the Free energy with respect to the shear strain tensor:
The expression for the shear modulus $C_\phi$ ($\phi=E,44,66$) is
\cite{Thal,Levy}
\begin{eqnarray}
C_{\phi}=C_{\phi,0} - \eta_\phi^2\chi_\phi ,
\label{eqn:Cx}
\end{eqnarray}
where $C_{\phi,0}=v_\phi^2\rho$ is the bare shear modulus,
where $v_\phi$ is the bare acoustic phonon velocity
and $\rho$ is the mass density.
$\eta_\phi$ is the quadrupole-strain coupling constant
due to the ``acoustic phonon'' given in Sec. \ref{sec:acoustic}.
In Eq. (\ref{eqn:Cx}), the condition for the structure transition, 
$C_\phi=0$, is satisfied when $\chi_\phi= g_\phi^{-1}(\gg1)$.
That is, the structure transition 
occurs prior to the divergence of $\chi_\phi$.

We can rewrite the expression for $C_{\phi}$ given in Eq. (\ref{eqn:Cx}) 
as follows:
\begin{eqnarray}
C_{\phi}^{-1}&=&C_{\phi,0}^{-1}[1 + g_\phi {\tilde \chi}_\phi]
\label{eqn:Cphi} ,\\
{\tilde \chi}_\phi&=& \chi_\phi/(1-g_\phi\chi_\phi)
\label{eqn:Cx2} ,
\end{eqnarray}
where $g_\phi\equiv \eta_\phi^2 C_{\phi,0}^{-1}$ is the effective el-el interaction
due to acoustic phonon given in the previous section.
In Eq. (\ref{eqn:Cphi}), the condition $C_\phi=0$
corresponds to the divergence of ${\tilde \chi}_\phi$,
since ${\tilde \chi}_\phi$ is the {\it total susceptibility} including the 
$e$-ph interactions due to acoustic phonons.

If we put $U=0$ for simplicity, Eq. (\ref{eqn:Cx2}) is expressed as
\begin{eqnarray}
{\tilde \chi}_{44}&=& \chi_{44}^{0}/(1-(g+g_{44})\chi_{44}^{0}),
 \label{eqn:chi44U0}\\
{\tilde \chi}_{E}&=& \chi_{E}^{0}/(1-(g+g_{E})\chi_{E}^{0}),
 \label{eqn:chiEU0}\\
{\tilde \chi}_{66}&=& \chi_{66}^{0}/(1-g_{66}\chi_{66}^{0}),
 \label{eqn:chi66U0}
\end{eqnarray}
where the suffix $0$ represents the bare susceptibility.
According to ${\tilde \chi}_{44}$ in Eq. (\ref{eqn:chi44U0}),
$g$ in Eq. (\ref{eqn:chiGU0}) for $\chi_{xz}^Q$ is replaced with $g+g_{44}$
when both optical and acoustic phonons are taken into account.
Therefore, we have to reduce $g$ to $g-g_{44}$
to keep the charge Stoner factor $\a_c$ and $\chi_{xz}^Q(\Q)$ unchanged.
Considering the relation $g_{44}\sim g_{E}$ that we derived 
in the previous section, we obtain ${\tilde \chi}_{E}\sim \chi_{E}$.
Then, we conclude that (i) $C_{44}\sim C_{44,0}$ since 
$\chi_{44}$ is seldom enhanced in the RPA.
Also, (ii) $C_{E}$ softens to some extent
since $\chi_{xy}^Q({\bm 0})$ is weakly enhanced
as shown in Fig. \ref{fig:chic-ac}, although
the relation $C_{E}=0$ will not be satisfied
because of the relation ${\tilde \chi}_{E}\sim \chi_{E}$.

As for $C_{66}$ given in Eq.(\ref{eqn:chi66U0}),
$\chi_{66}^{0}$ in the present model is $\sim 2 {\rm eV}^{-1}$,
while we estimate $g_{66}=0.1\sim0.2{\rm eV}$.
Therefore, we expect (iii) $C_{66}\sim C_{66,0}$ in the RPA,
which is inconsistent with experimentally observed
large softening in $C_{66}$ \cite{Yoshizawa}.
In the RPA, the softening in shear modulus ($C_{66}$, $C_{44}$ and $C_E$)
is small according to Eqs. (\ref{eqn:chiE-def})-(\ref{eqn:chi66-def})
and Figs. \ref{fig:chic} (a)-(c).
In the next subsection, we analyze $\chi_{66}$ 
by taking account of the two-orbiton process
that is not included in the RPA.


\subsection{Softening due to two-orbiton process;
the Aslamazov-Larkin-type diagram
\label{sec:AL}}

In the previous subsection,
we studied the softening of shear moduli within the RPA.
However, the obtained softening is very small
since only the AFQ fluctuations develop in the present model.
In this subsection, we analyze $\chi_{66}$ 
by taking account of the ``two-orbiton process''
that is not included in the RPA.
In usual cases, this higher order process is negligible.
However, it gives divergent increase in $\chi_{66}$,
and creates the FQ-QCP near the AFQ-QCP.
For this reason, the orthorhombic structure transition ($C_{66}=0$)
is induced by the two-orbiton process.

Before calculating the two-orbiton process,
let us consider the third order anharmonic phonon-phonon
coupling $\sim A_{(3)}u_x^2u_{66}$, where $u_{x}$ is the 
displacement of Fe-ion optical mode in Eq. (\ref{eqn:e-ph}), 
and $u_{66}$ is the displacement of As-ion acoustic node in 
Eq. (\ref{eqn:V66}) or (\ref{eqn:V66O}).
Figure \ref{fig:AL} (b) shows the
second-order-term with respect to $A_{(3)}$,
where $D(q)$ is the optical phonon Green function in Eq. (\ref{eqn:D}),
and the factor $1/2$ is introduced to cancel the overcounting 
with respect to the upside-down diagrams.
This term gives a self-energy correction for the acoustic phonon 
Green function.
This two phonon process would be measurable in Raman spectroscopy.
By considering the $e$-ph interaction, 
each $D(q)$ in (b) is replaced with $D(q)+(\eta D(q))^2\cdot\chi_{yz}^Q(q)$.

Here, we study the ferro-quadrupole susceptibility due to
the ``two-orbiton term'' given by Fig. \ref{fig:AL} (c), 
and analyze for the softening of $C_{66}$.
Since the coefficient of the anharmonic phonon-phonon coupling $A_{(3)}$
would be small in iron pnictides,
we instead consider the three-point vertex given by 
electron Green functions in Fig. \ref{fig:AL} (d).
The two-orbiton term in Fig. \ref{fig:AL} (c) is similar to the
Aslamazov-Larkin (AL) term for the excess conductivities
due to superconducting fluctuations;
such as longitudinal \cite{AL} and Hall \cite{Fuku} conductivities,
and the Nernst coefficient \cite{Uss}.
Figure \ref{fig:AL} (c) is a virtual process in which
one $O_{x^2-y^2}$-type orbiton with $\q=0$ breaks into 
two $O_{xz}$-type orbitons with zero total momentum ($\pm \Q$).
After taking the momentum summation, the two-orbiton term (c) 
in 2D systems would be strongly enhanced for $\q=0$ since
$\chi_{yz}^Q(\k)$ has a large peak at finite momentum $\k$. 
The mathematical expression for the two-orbiton process in 
Fig. \ref{fig:AL} (c) for $\chi_{66}$, which we call 
$\chi_{66}^{\rm TO}$, is given as
\begin{eqnarray}
\chi_{66}^{\rm TO}&=& \frac12 (2g)^{2}T\sum_{\q,l,\Gamma,\Gamma'} 
A_{x^2-y^2}(\Gamma,\Gamma';\q,\w_l)A_{x^2-y^2}(\Gamma,\Gamma';\q,\w_l)
 \nonumber\\
& &\times (1+2g\chi_{\Gamma}^{Q}(\q,\w_l))(1+2g\chi_{\Gamma'}^{Q}(\q,\w_l))
 \nonumber\\
&\approx& \frac12(2g)^{4}T\sum_{\q,l,\Gamma,\Gamma'} 
A_{x^2-y^2}(\Gamma,\Gamma';\q,\w_l)A_{x^2-y^2}(\Gamma,\Gamma';\q,\w_l)
 \nonumber\\
& &\times \chi_{\Gamma}^{Q}(\q,\w_l)\chi_{\Gamma'}^{Q}(\q,\w_l)
\label{eqn:chi66} ,
\end{eqnarray}
where $A_{x^2-y^2}(\Gamma,\Gamma';\q,\w_l)$
is the three point vertex with respect to ${\hat V}_{66}$
in Eq. (\ref{eqn:V66O}) and quadrupole operators
${\hat O}_{\Gamma}$ and ${\hat O}_{\Gamma'}$, shown in Fig. \ref{fig:AL} (d).
We used the relation $\chi_\Gamma^Q(q)=\chi_\Gamma^Q(-q)$.
When $U=0$, $A_{x^2-y^2}$ for $\w_l=0$ is given as
\begin{eqnarray}
A_{x^2-y^2}(\Gamma,\Gamma';\q) &=& -2T\sum_{n,\k}{\rm Tr}\left\{ 
{\hat G}_\k(\e_n){\hat o}_{x^2-y^2} {\hat G}_\k(\e_n) \right.
 \nonumber \\
& &\left. \times {\hat o}_\Gamma{\hat G}_{\k+\q}(\e_n){\hat o}_{\Gamma'} \right\}
 \label{eqn:Aw0} ,
\end{eqnarray}
where the factor 2 in front of Eq. (\ref{eqn:Aw0})
accounts for the diagrams with reversing three Green functions
${\hat G}_\k(\e_n)\rightarrow {\hat G}_{-\k}(-\e_n)$.
Near the QCP $g\lesssim g_c$,
the most divergent quadrupole susceptibility is $\chi_{xz(yz)}^{Q}$.
Therefore, the dominant contribution for $\chi_{66}^{\rm TO}$ 
in Eq. (\ref{eqn:chi66}) will be given by the term
with $\Gamma=\Gamma'=xz$ or $yz$.
After the analytic continuation, the functional form of 
$\chi_{xz/yz}^{Q}(\q,\w)$ for $\q\approx {\bm Q}_{xz}=(\pi,0)$ 
or $\q\approx {\bm Q}_{yz}=(0,\pi)$ would be approximately given as
\begin{eqnarray}
\chi_\Gamma^{Q}(\q,\w+i\delta)&=&\frac{c\xi^2}
{1+ \xi^2(\q-\Q_\Gamma)^2-i\w/\w_0},
 \label{eqn:chi-analytic}
\end{eqnarray}
for $\Gamma=xz$ or $yz$,
where $\xi$ is the correlation length
and $\w_0$ is the characteristic energy of the fluctuation.
The relation $\xi^2\propto \w_0^{-1}$ holds in the RPA.

Next, we consider the temperature dependence of $\xi$.
In the FLEX approximation \cite{Onari} or SCR theory \cite{Moriya},
the bare susceptibility $\chi_\phi^0$ is approximately suppressed as
$\chi_\phi^0-\a T$ ($\a>0$)
due to the thermal fluctuations, which are described as
the self-energy and Maki-Thompson vertex corrections.
In this case, we obtain
$\chi_{xz}^{Q}({\bm Q},0)\propto (1-g\chi_{xz}^0({\bm Q},0)+g\a T)^{-1}
\propto (T-T_{\rm AFQ})^{-1}$ based on the RPA,
where $T_{\rm AFQ}=-(1-g\chi_{xz}^0({\bm Q},0))/g\a$
is the transition temperature to the AFQ ordered state.
Since $\chi_{xz}^{Q}({\bm Q},0)\propto\xi^2$,
we assume the following relations
\begin{eqnarray}
\xi^2&=&l(T-T_{\rm AFQ})^{-1},
 \\
\w_0&=&l'(T-T_{\rm AFQ}),
 \label{eqn:w0}
\end{eqnarray}
where $l,l'$ are constants.
Note that $\w_0\xi^2$ is temperature independent \cite{Moriya}. 
By carrier doping, $T_{\rm AFQ}$ changes from positive to negative, while 
other model parameters ($c$, $l$, and $l'$) would be insensitive to doping.
As shown in Fig. \ref{fig:chic-ac}, 
$\chi_{xz}^{Q}(\Q,0)\sim 2.4\times(1-\a_c)^{-1}\sim 12\xi^2$.
In the case (i) $T_{\rm AFQ}>0$, the relation $\w_0<T$ 
is satisfied near $T_{\rm AFQ}$.
In the opposite case (ii) $T_{\rm AFQ}<0$, the relation $\w_0>T$ 
will hold for wide range of temperatures.
Note that the present phenomenological model 
in Eqs. (\ref{eqn:chi-analytic})-(\ref{eqn:w0})
is reproduced by the microscopic calculation by FLEX approximation
\cite{Onari}.
As for the spin propagator in cuprate superconductors,
the relation $\w_0>T$ ($\w_0<T$) holds in over-doped (under-doped) systems.

Here, we comment on the self-energy correction 
and the Maki-Thompson-type vertex correction for $\chi_{66}^{\rm TO}$,
shown in Figs. \ref{fig:AL} (e) and (f) respectively.
The former term is included in the FLEX approximation,
and it gives the Curie-Weiss behavior of $\chi^Q_{xz}({\bm Q},0)$
given by Eqs. (\ref{eqn:chi-analytic})-(\ref{eqn:w0}),
as reported in Ref. \onlinecite{Onari} or Ref. \onlinecite{Moriya}.
The latter term would be negligible since its temperature dependence 
is smaller than that of the former term.
For this reason, we concentrate on the two-orbiton term
in Fig. \ref{fig:AL} (c) hereafter.

\begin{figure}[!htb]
\includegraphics[width=0.7\linewidth]{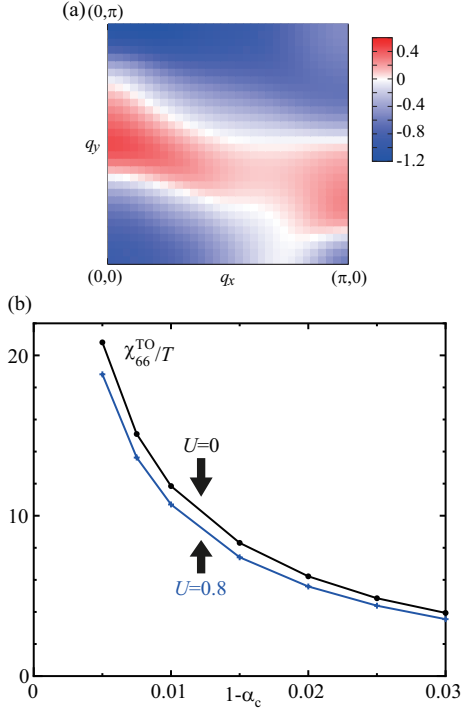}
\caption{
(Color online) 
(a) Obtained $A_{x^2-y^2}(xz,xz;\q)$ for $U=0$.
We put $n=6.05$ and $T=0.05$.
(b) $\chi_{66}^{\rm TO}/T$ given by Eq. (\ref{eqn:chi66-highT})
for $U=0$ and $U=0.8$ as function of $\a_c$.
Using the relation $6\xi^2\sim(1-\a_c)^{-1}$ given 
in the caption of Fig. \ref{fig:chic}, we obtain 
$\chi_{66}^{\rm TO}/T\sim0.1(1-\a_c)^{-1}\sim 0.6\xi^2$.
}
\label{fig:quad}
\end{figure}

From now on, we perform the numerical calculation of the two-orbiton process 
in the case (i), in which the relations $\xi\gg1$ and $\w_0\ll T$
are realized near the orbital-ordered state.
In this case, the dominant contribution in Eq. (\ref{eqn:chi66})
comes form the terms with $\Gamma=\Gamma'=xz$ and $yz$.
Also, we can safely apply the classical approximation, 
in which the terms with $\w_l\ne0$ are dropped in Eq. (\ref{eqn:chi66}).
Under these approximations, Eq. (\ref{eqn:chi66}) is simplified as
\begin{eqnarray}
\chi_{66}^{\rm TO} &=& (2g)^{4}T\sum_\q 
\{ A_{x^2-y^2}(xz,xz;\q) \chi_{xz}^{Q}(\q,0) \}^2 .
\label{eqn:chi66-highT}
\end{eqnarray}
To calculate $A_{x^2-y^2}(\Gamma,\Gamma';\q)$,
we introduce a uniform FQ potential term
$H'=h\sum_i {\hat O}_{x^2-y^2}^i$, where $h$ is an infinitesimal small constant.
Then, the three point vertex is given as
the following Ward identity \cite{AGD}:
\begin{eqnarray}
A_{x^2-y^2}(\Gamma,\Gamma';\q) =
\frac1h \left[ {\bar \chi}_{\Gamma,\Gamma'}^{Q}(\q,0;h)
-{\bar \chi}_{\Gamma,\Gamma'}^{Q}(\q,0;0) \right]
 \label{eqn:Ax2y2-W} ,
\end{eqnarray}
where ${\bar \chi}_{\Gamma,\Gamma'}^{Q}(\q,\w_l;h)$ is the 
``irreducible'' quadrupole susceptibility with respect to $g$.
In the numerical calculation, we have to fix $\mu$ against the change in $h$.
Equation (\ref{eqn:Ax2y2-W}) gives the correct
three-point vertex even for $U\ne0$.
In the case of $U=0$, ${\bar \chi}_{\Gamma,\Gamma'}^{Q}$ is simply
given as ${\bar \chi}_{\Gamma,\Gamma'}^{Q}(\q,\w_l;h)
=-T\sum_{\k,n}{\rm Tr}\{ {\hat o}_\Gamma {\hat G}(\k+\q,\e_n+\w_l;h)
{\hat o}_{\Gamma'} {\hat G}(\k,\e_n;h)\}$.
The obtained $A_{x^2-y^2}(xz,xz;\q)$ for $U=0$ is presented
in Fig. \ref{fig:quad} (a).
Similar result is obtained for $U=0.8$eV.

According to the functional form of $\chi_{xz}^{Q}(\q,0)$
in Eq. (\ref{eqn:chi-analytic}),
$\chi_{66}^{\rm TO}/T\propto \sum_\q \{ \chi_{xz}^{Q}(\q,0) \}^2
\propto\xi^2 \propto(T-T_{\rm AFQ})^{-1}$ in two-dimensional systems.
The numerical result for $\chi_{66}^{\rm TO}/T$ given in 
Eq. (\ref{eqn:chi66-highT}) is shown in Fig. \ref{fig:quad} (b).
The obtained result follows the relation
$\chi_{66}^{\rm TO}/T \sim 0.1(1-\a_c)^{-1}$.
Since $6\xi^2\sim (1-\a_c)^{-1}$,
the relation $\chi_{66}^{\rm TO}/T\sim 0.6\xi^2$ is verified numerically.



In the same way, two-orbiton processes for other two 
shear modulus susceptibilities, $\chi_{44}^{\rm TO}$ and $\chi_{E}^{\rm TO}$,
are proportional to the square of 
$A_{\Gamma}(xz,xz;\q)$ for $\Gamma=xz$ and $xy$, respectively.
However, they are four orders of magnitude smaller than 
$\chi_{66}^{\rm TO}$, as recognized in TABLE \ref{tab:tab1}:
This selection rule for $A_{\Gamma}$ is understood as follows:
According to the relation 
${\rm Tr}\{ {\hat O}_{\Gamma}{\hat O}_{xz}{\hat O}_{xz} \}=0$ 
for $\Gamma=xz,xy$, we recognize that $A_{xz/xy}$ originates from the 
off-diagonal terms of the Green function $G_{l,m}$ ($l\ne m$)
that is much smaller than the diagonal terms.
For this reason, the two-orbiton process 
is negligible except for $\chi_{66}^{\rm TO}$.


\begin{table}[h]
\begin{center}
\begin{tabular}{c|c|c|c|c|c}   
\hline
$\Gamma$ & & $x^2-y^2$ & $xz$ & $yz$ & $xy$ \\
\hline
$A_\Gamma(xz,xz,{\bm Q})$ & & $-0.60$ & 
$1.9\times10^{-3}$ & $-1.0\times10^{-3}$ & $3.2\times10^{-4}$ \\
\hline
\end{tabular}
\caption{\label{tab:tab1}
Three point vertex $A_\Gamma(xz,xz,{\bm Q})$ for 
$\Gamma=x^2-y^2$, $xz$, $yz$, and $xy$.
${\bm Q}=(\pi,0)$ corresponds to the peak position of
$\chi_{xz}^Q(\q,0)$ in the five-orbital model.
We recognize that $A_{\Gamma}(xz,xz,{\bm Q})\sim O(1)$ only for $\Gamma=x^2-y^2$;
This selection rule means that $\chi_{44,E}^{\rm TO}\ll1$.
}
\end{center}
\end{table}

Here, we discuss the softening of $C_{66}$
by taking the two-orbiton process into account:
According to Eqs. (\ref{eqn:Cphi}) and (\ref{eqn:Cx2}), we obtain
\begin{eqnarray}
C_{66}^{-1}&=&C_{66,0}^{-1}[1+g_{66}{\tilde \chi}_{66}]
\label{eqn:Cx66} ,\\
{\tilde \chi}_{66}&=& \frac{a_{66}+\chi_{66}^{\rm TO}}
{1-g_{66}(a_{66}+\chi_{66}^{\rm TO})}
\label{eqn:tildeC66} ,
\end{eqnarray}
where $a_{66}\equiv 2\chi_{66}^{0}({\bm 0},0)$.
Now, we consider the case (i) $T_{\rm AFQ}>0$ and $\w_0\ll T$.
As we have obtained the relation $\chi_{66}^{\rm TO}\propto T\xi^2$,
we put $\chi_{66}^{\rm TO}= b_{66}T/(T-T_{\rm AFQ})$.
Since the temperature dependence of $a_{66}$ is small, we obtain
\begin{eqnarray}
{\tilde \chi}_{66}&=&\frac{a_{66}+b_{66}}{1-g_{66}(a_{66}+b_{66})}
\frac{T-(a_{66}/(a_{66}+b_{66}))T_{\rm AFQ}}{T-T_S} ,
 \nonumber \\
 \label{eqn:tildeC66-2} \\
T_S&=& T_{\rm AFQ}\frac{1-g_{66}a_{66}}{1-g_{66}(a_{66}+b_{66})} \ \ (>T_{\rm AFQ})
 \label{eqn:TS} .
\end{eqnarray}
Then, the difference between $T_S$ and $T_{\rm AFQ}$,
which is conventionally denoted as $E_{\rm JT}$ \cite{Thal,Levy},
is given by $E_{\rm JT}=T_S(g_{66}b_{66})/(1-g_{66}a_{66})>0$.
According to Eq. (\ref{eqn:tildeC66-2}),
Eq. (\ref{eqn:Cx66}) is rewritten as
\begin{eqnarray}
 C_{66} = C_{66,0}(1-g_{66}(a_{66}+b_{66}))\frac{T-T_S}{T-T_{\rm AFQ}},
 \label{eqn:C66-temp}
\end{eqnarray}
Here, $g_{66}=0.1\sim0.2$eV and $a_{66}\sim2{\rm eV}^{-1}$.
We stress that Eqs. (\ref{eqn:tildeC66-2})-(\ref{eqn:C66-temp})
are valid only for $\w_0\ll T$.



Finally, we calculate the two-orbiton process
analytically for general value of $\w_0/T$.
Since we cannot apply the classical approximation that 
was used to derive Eq. (\ref{eqn:chi66-highT}),
we have to perform the analytic continuation \cite{AGD}
of Eq. (\ref{eqn:chi66}).
The obtained expression including the quantum fluctuation
contribution is given as 
\begin{eqnarray}
\chi_{66}^{\rm TO}&=& (2g)^{4}\{A_{x^2-y^2}(xz,xz;{\bm Q})\}^2
\sum_{\q}\int_{-\infty}^\infty \frac{dx}{\pi} n(x)
\nonumber \\
& &\times 2{\rm Im}\chi_{xz}^Q(\q,x+i\delta)
{\rm Re}\chi_{xz}^Q(\q,x+i\delta)
\label{eqn:chi66-analytic} ,
\end{eqnarray}
where $n(x)=(e^{\beta x}-1)^{-1}$ is the Bose distribution function,
and we put $A_{x^2-y^2}$ outside of the $\q$-summation
since its momentum dependence is much smaller than
that of $\chi_{xz}^Q$.
First, we perform the $x$-integration using the following equations:
\begin{eqnarray}
&& \frac1{e^x-1}= \sum_{n=1}^\infty \frac{2x}{(2n\pi)^2+x^2}+\frac1x , 
 \\
&& \int_{-\infty}^\infty \frac{x^2}{(a^2+x^2)(b^2+x^2)^2}dx
=\frac{\pi}{2b(a+b)^2} ,
\end{eqnarray}
where $a, b>0$.
Then, the expression after the $x$-integration 
in Eq. (\ref{eqn:chi66-analytic}) is given as
\begin{eqnarray}
c^2\xi^4\left( \w_0^2T\sum_{n=1}^\infty (B_\q\w_0+2n\pi T)^{-2}
+ \frac{T}{2B_\q^2} \right),
\label{eqn:chi66-3}
\end{eqnarray}
where $B_\q=1+\xi^2(\q-{\bm Q})^2$.
Next, we take the $\q$-summation 
$\sum_\q \approx \frac1{2\pi}\int_0^{\pi}qdq$
under the assumption $\xi^2\gg1$.
Then, the $\q$-summation of the second term in Eq. (\ref{eqn:chi66-3}) 
is easily obtained as $c^2\xi^2T/8$.
Also, the $\q$-summation of the first term is given as
\begin{eqnarray}
&&\frac{c^2\xi^2\w_0}{8\pi}\sum_{n=1}^{n_{\rm max}}\frac1{n+\frac{\w_0}{2\pi T}}
\nonumber \\
&&=\frac{c^2\xi^2\w_0}{8\pi}\left[ \psi\left(n_{\rm max}+\frac{\w_0}{2\pi T}
+1\right) -\psi\left(\frac{\w_0}{2\pi T}+1\right) \right] ,
 \nonumber \\
\end{eqnarray}
where $\psi(x)$ is di-Gamma function, and
the cutoff $n_{\rm max}\equiv(1+\xi^2\pi^2)\w_0/2\pi T$ originates from 
the fact that the $\q$-summation is limited to the region $|\q|\le\pi$
in periodic systems.
As a result, the final expression for the two-orbiton term is 
\begin{widetext}
\begin{eqnarray}
\chi_{66}^{\rm TO} &=& X \xi^2
\left\{ \frac{\w_0}{\pi} \left[ \psi\left(n_{\rm max}+\frac{\w_0}{2\pi T}
+1\right) -\psi\left(\frac{\w_0}{2\pi T}+1\right) \right] +T\right\}
\label{eqn:chi66-analytic2} ,
\end{eqnarray}
\end{widetext}
where $\displaystyle X\equiv 
\frac{(2g)^4 c^2}{4\pi}\{A_{x^2-y^2}(xz,xz;{\bm Q})\}^2$.
Here, we verify Eq. (\ref{eqn:chi66-analytic2}) in the 
opposite two limits:
In the case (i) $\w_0\ll T$,
the di-Gamma functions in Eq. (\ref{eqn:chi66-analytic2}) are negligible.
By applying the relation $\xi^2=l/(T-T_{\rm AFQ})$, we obtain
\begin{eqnarray}
\chi_{66}^{\rm TO} &\approx& X \xi^2 T
\nonumber \\
&\approx& b\frac{T}{T-T_{\rm AFQ}} 
\label{eqn:chi66-analytic-highT} ,
\end{eqnarray}
where $b=Xl$.
The first line in Eq. (\ref{eqn:chi66-analytic-highT})
coincides with Eq. (\ref{eqn:chi66-highT})
since $\sum_\q\{\chi_{xz}^Q(\q,0)\}^2 = c^2\xi^2/4\pi$.
In the opposite case (ii) $\w_0\gg T$, the term $T$ in the curly brackets 
in Eq. (\ref{eqn:chi66-analytic2}) is negligible.
Taking the relations $\psi(x)\approx \log(x)$ for $x\gg1$ and
$\w_0\xi^2\propto\xi^0$ into account, we obtain
\begin{eqnarray}
\chi_{66}^{\rm TO} &\approx& X \xi^2\w_0 \log(2+\pi^2\xi^2)
\nonumber\\
&\approx& b'_{66}\log\left(\frac{\pi^2 l}{T-T_{\rm AFQ}}\right)
\label{eqn:chi66-analytic-lowT} ,
\end{eqnarray}
where $b'=X \xi^2\w_0$.
Therefore, in the case (ii) $T_{\rm AFQ}<0$ and $\w_0\gg T$, 
${\tilde \chi}_{66}$ in Eq. (\ref{eqn:Cx66}) is given 
by replacing $\chi_{66}^{\rm TO}$ with $b'_{66}\log({\pi^2 l}/(T-T_{\rm AFQ}))$ 
in Eq. (\ref{eqn:tildeC66}).
In this case, the temperature dependence of 
$\chi_{66}^{\rm TO}$ is much moderate.
In Sec. \ref{sec:Dis-C66}, we will discuss the temperature dependence 
of $C_{66}$ based on Eq. (\ref{eqn:chi66-analytic2}).

In the above derivation, we have neglected the effect of
mass-enhancement factor brought by the third point vertex.
If we take this effect into account, both
Eqs. (\ref{eqn:chi66-analytic-highT}) and (\ref{eqn:chi66-analytic-lowT})
are multiplied by the factor $(m^*/m)^2=2^2\sim3^2$,
as we will discuss in Sec. \ref{sec:Dis-C66}.

\section{Discussions
\label{sec:discussion}}


\subsection{Why $O_{x^2-y^2}$-FQ fluctuations are the most 
divergent for $T_{\rm AFQ}>0$?}

In this paper, we have studied the development of 
quadrupole susceptibilities in iron pnictides
based on the RPA and beyond the RPA.
The main fluctuations in the present study is the $O_{xz/yz}$-AFQ
fluctuations, $\chi_{xz/yz}^Q({\bm Q})$,
which are produced by Fe-ion in-plane optical phonons.
The acoustic phonons $\q\sim\Q$ with finite energy also
assist in producing the AFQ fluctuations; see Eq. (\ref{eqn:chi44U0}).
We also find that the $O_{x^2-y^2}$-FQ fluctuations, $\chi_{66}$,
are induced by the ``two-orbiton process'' described by 
the AL-type diagram in Fig. \ref{fig:AL} (c).
Here, the anharmonic three-phonon coupling 
is produced by the three-point vertex in Eq. (\ref{eqn:Aw0}).
The two-orbiton process is important in iron pnictides
because of the two-dimensionality.



We discuss on the {\it total susceptibility} by including
the electron-acoustic phonon interaction, ${\tilde \chi}_\Gamma^Q(\q)$,
by taking the two-orbiton process into account.
For $\Gamma=xz$ and $x^2-y^2$,
\begin{eqnarray}
{\tilde \chi}_{xz}^Q({\bm Q})
 &=& \frac{\chi_{xz}^0({\bm Q})}{1-(g+g_{44})\chi_{xz}^0({\bm Q})},
 \\
{\tilde \chi}_{x^2-y^2}({\bm 0})
 &=& \frac{\chi_{66}^0({\bm 0})+\chi_{66}^{\rm TO}}
 {1-g_{66}(\chi_{66}^0({\bm 0})+\chi_{66}^{\rm TO})},
\end{eqnarray}
where $\chi_{66}^{\rm TO}$ is proportional to the square of the AFQ
correlation length $\xi^2\propto {\tilde \chi}_{xz}^Q({\bm Q})$.
When $g_{66}=0$, therefore,
${\tilde \chi}_{xz}^Q({\bm Q})$ and 
${\tilde \chi}_{x^2-y^2}^Q({\bm 0})$
diverges at the same time in proportion to $\xi^2$.
As $g_{66}$ increases from zero, only ${\tilde \chi}_{x^2-y^2}^Q({\bm0})$
is enhanced because of the absence of the two-orbiton process;
$A_\Gamma(xz,xz;\q),A_\Gamma(yz,yz;\q)\ll1$ for $\Gamma=xz$ or $yz$
as shown in TABLE \ref{tab:tab1}.
For this reason, the relation $T_S>T_{\rm AFQ}$ is universally satisfied
even if a fully self-consistent calculation is performed.

As results, both the AFQ fluctuations (=origin of high-$T_{\rm c}$)
and the FQ fluctuations (=origin of shear modulus softening)
develop at the same time, and latter fluctuations overcome
the former near the $T_S$.
The orthorhombic phase transition 
in under-doped compounds is brought by the divergence of
the two-orbiton process ${\tilde \chi}_{66}$.


\subsection{Softening of $C_{66}$: comparison between theory and experiment
\label{sec:Dis-C66}}

Is this subsection, we discuss the softening of $C_{66}$
in under- and over-doped iron pnictides
based on the results in Sec. \ref{sec:shear}.
For this purpose, we first estimate the magnitude of the three-point vertex
$A_{x^2-y^2}(xz,xz,\Q)$ based on the Ward identity given 
in Eq. (\ref{eqn:Ax2y2-W}).
Since ${\bar \chi}_{xz,xz}^{Q}(\Q,0;0)\sim(2g)^{-1}$,
$|A_{x^2-y^2}|\sim (2g)^{-1}/\delta E$, where $\delta E$ is the 
bandwidth of the $xz/yz$ band.
Since $2g\sim0.5{\rm eV}$ and $\delta E\sim 2{\rm eV}$ 
according to the band calculations \cite{Kuroki},
and considering the effect of band renormalization due to the 
mass-enhancement factor $m^*/m\ (=2\sim3)$ \cite{Shishido},
we expect $|A_{x^2-y^2}|\sim 1(m^*/m)\ [{\rm eV}^{-2}]$.
This rough estimation is consistent with the numerical result
in Fig. \ref{fig:quad} (a).

Now, we discus the under-doped case with $T_{\rm AFQ}>0$.
In this case, $\chi_{66}^{\rm TO}/T\approx X\xi^2$
shown in Eq. (\ref{eqn:chi66-analytic-highT}).
Using the relations $c\sim12$ and $6\xi^2\sim (1-\a_c)^{-1}$
as discussed in the caption of Fig \ref{fig:chic}, we obtain 
$X\sim 0.7$ and $\chi_{66}^{\rm TO} \sim 0.12(m^*/m)^2(1-\a_c)^{-1}$.
This estimation is consistent with the numerical result
Fig. \ref{fig:quad} (b) if we put $(m^*/m)=1$.

We also discuss the optimum or over-doped systems without 
structure transition, in which the relation $\w_0\gg T$ is satisfied.
In this case, $\chi_{66}^{\rm TO}\sim b'\log(\pi^2l/(T-T_{\rm AFQ}))$.
Since the temperature dependence of $\chi_{66}^{\rm TO}$ is moderate,
$\chi_{66}^{\rm TO}$ would be comparable or smaller than $a_{66}$.
Therefore, in over-doped systems,
the softening in $C_{66}$ would be much moderate, showing a 
deviation from the Curie-Weiss type form in Eq .(\ref{eqn:C66-temp}).

\begin{figure}[!htb]
\includegraphics[width=0.98\linewidth]{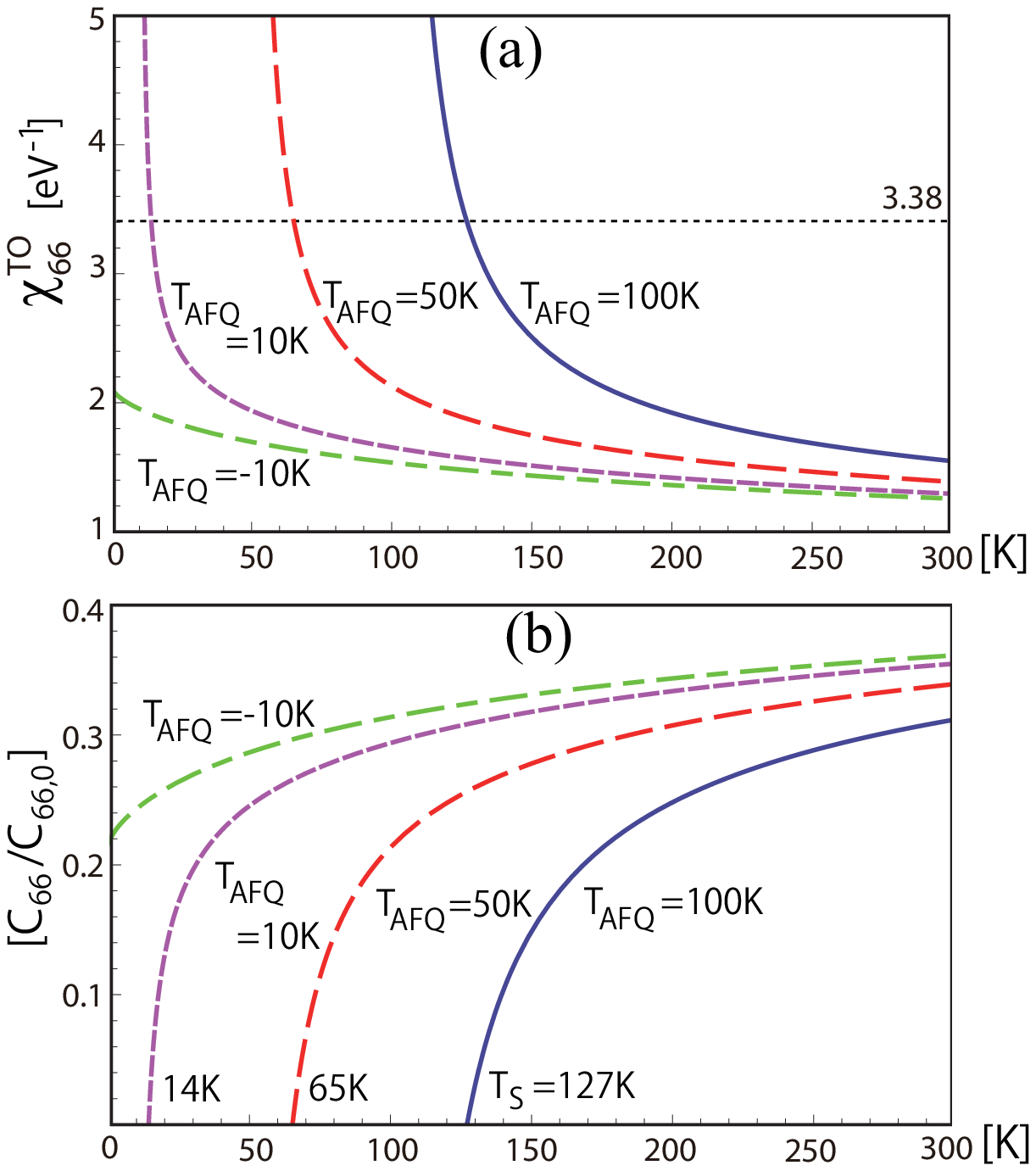}
\caption{
(Color online) 
(a) $\chi_{66}^{\rm TO}$ for $T_{\rm AFQ}=100$K, $50$K, $10$K and $-10$K.
given by Eq. (\ref{eqn:chi66-analytic2}).
(b) $C_{66}/C_{66,0}$ for $T_{\rm AFQ}=100$K, $50$K, $10$K and $-10$K.
given by Eqs. (\ref{eqn:Cx66}) and (\ref{eqn:tildeC66}).
We put $X=5.7$, $l=0.086$, $l'=2$, $g_{66}=0.17$, and $a_{66}=2.5$.
$C_{66}=0$ is realized when $\chi_{66}^{\rm TO}=g_{66}^{-1}-a_{66}$,
which is $3.38$ in the present parameters.
Using these {\it same} parameters, we can fit the recent 
experimental data by Yoshizawa {\it et al.} \cite{comment4}
for Ba(Fe$_{1-x}$Co$_x$)$_2$As$_2$ for $x=0\sim0.1$,
just by changing $T_{\rm AFQ}$.
}
\label{fig:MATH}
\end{figure}

Now, we analyze the temperature dependence of $\chi_{66}^{\rm TO}$
and $C_{66}/C_{66,0}$ by using Eq. (\ref{eqn:chi66-analytic2}).
We can fix the prefactor 
$(2g)^4c^2\xi^2\{A_{x^2-y^2}\}^2/4\pi \equiv X \xi^2$ 
in front of Eq. (\ref{eqn:chi66-analytic2})
based on the relation $\chi_{66}^{\rm TO}/T = X\xi^2$ for $\w_0\ll T$:
We obtain $X\sim 0.6$ according to Fig. \ref{fig:quad} (b).
Hereafter, we put $X=5.4$ by multiplying the 
square of the mass-enhancement factor, ($m^*/m)^2\sim9$.
We also put $\w_0=l'(T-T_{\rm AFQ})$ with $l'=2$,
and $\xi^2=l(T-T_{\rm AFQ})^{-1}$ with $l=0.086\ [{\rm eV}]$, 
which means that $\xi\sim2$ when $T-T_{\rm AFQ}=250$K.
Using the obtained $\chi_{66}^{\rm TO}$,
we plot $C_{66}/C_{66,0}$ in Fig. \ref{fig:MATH} (b)
based on Eqs. (\ref{eqn:Cx66}) and (\ref{eqn:tildeC66}).
Here, we set $g_{66}=0.17$eV and $a_{66}=2.5{\rm eV}^{-1}$.
In the case of $T_{\rm AFQ}=100$K, we obtain $E_{\rm JT}\approx27$K. 
In the FLEX approximation \cite{Onari},
$T_{\rm AFQ}$ changes from positive to negative by carrier doping, while 
other parameters ($X$, $l$, $l'$ and $a_{66}$) are insensitive to the doping.
Similarly to Fig. \ref{fig:MATH} (b), we can fit the recent experimental 
data by Yoshizawa {\it et al.} \cite{comment4}
for Ba(Fe$_{1-x}$Co$_x$)$_2$As$_2$ with $x=0\sim0.1$, by choosing $T_{\rm AFQ}$ 
while other parameters ($X$, $l$, $l'$ and $a_{66}$) are fixed.
This fact is a strong evidence for the success of orbital fluctuation theory
in iron pnictide superconductors.

In the present paper, we consider that the origin of high-$T_{\rm c}$
is the AFQ fluctuations.
On the other hand, Yanagi {\it et al}. \cite{Ohno2} claimed that 
high-$T_{\rm c}$ originates from the FQ fluctuations 
that give the softening in $C_{66}$:
In the latter mechanism, a rough estimation of $T_{\rm c}$ is given as
\begin{eqnarray}
T_{\rm c}\sim \w_c \exp(-(1+\beta\lambda)/\beta\lambda),
\label{eqn:BCS}
\end{eqnarray}
where $\w_c$ is 
the phonon energy relevant for the orbital fluctuations,
which is just $\sim10$K for $|\q|\sim0.1\pi$.
$\beta\equiv 1+g_{66}{\tilde \chi}_{66}=C_{66,0}/C_{66}$ is the 
enhancement factor due to FQ fluctuations \cite{comment}.
However, $C_{66,0}/C_{66}$ observed in optimally-doped Ba(Fe,Co)$_2$As$_2$
is just $\sim1.2$ \cite{Fernandes,comment4}:
Apparently, such small enhancement cannot reproduce
high-$T_{\rm c}$ superconductivity in iron-pnictides.




In the present study, in contrast, weak softening in optimally-doped 
sample is ascribed to the change in the scaling of $\chi_{66}^{\rm TO}$,
not to the weakness of AFQ fluctuations.
In fact, the softening is moderate in the case of $T_{\rm AFQ}=-10$K
in Fig. \ref{fig:MATH}, while the AFQ correlation 
$\xi^2\approx 1000/(T{\rm [K]}+10)$ is enough to cause the 
superconductivity at $T_{\rm c}\sim30$K.
Therefore, moderate softening and high-$T_{\rm c}$ are compatible
in the present study.





\subsection{Quadrupole-ordered state in under-doped compounds}

Here, we consider the orbital or quadrupole ordered state in 
under-doped compounds.
In the mean-field approximation for the multiorbital Hubbard model 
for iron pnictides \cite{OO-theory-MF,OO-theory-MF2},
stripe-type SDW order occurs for $U>U_{\rm c}$,
and weak orbital polarization ($n_{xz}\ne n_{yz}$) is induced 
as the secondary order when the magnetization is large.
However, in real materials, orthorhombic transition occurs
in the paramagnetic state, and the SDW-order is induced
in the orthorhombic phase.
To solve this problem, 
we studied the multiorbital HH model beyond the mean-field theory,
and found that the FQ order occurs in the paramagnetic state
due to the two-orbiton process.
Fortunately, this FQ order does produce the experimentally observed 
SDW order, as we will explain in the next subsection.

\begin{figure}[!htb]
\includegraphics[width=0.9\linewidth]{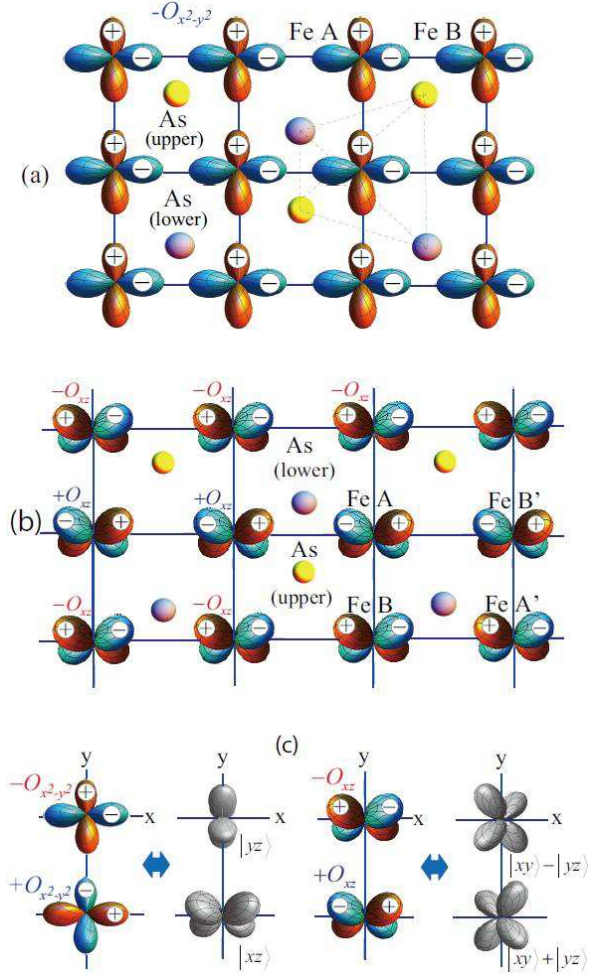}
\caption{
(Color online) 
(a) $O_{x^2-y^2}$-FQ order given by the divergence of $C_{66}^{-1}$.
Be careful not to confuse ${\hat O}_{x^2-y^2}$ with the 
$x^2-y^2$-orbital operator.
(b) $O_{xz}$-AFQ order brought by the divergence of $\chi_{xz}^Q({\bm Q})$.
(c) The correspondence between $O_{x^2-y^2}$-quadrupole order 
($O_{xz}$-quadrupole order) and the $d$-orbital 
with larger occupation number.
}
\label{fig:OO-fig}
\end{figure}

As discussed in Sec. \ref{sec:AL}, the divergence of ${\tilde \chi}_{66}$,
which is the total FQ susceptibility 
given by both optical and acoustic phonons,
causes the orthorhombic structure transition when $C_{66}=0$.
The $O_{x^2-y^2}$-FQ order is realized in the orthorhombic phase.
The schematic quadrupole order is shown in Fig. \ref{fig:OO-fig} (a).
Since $O_{x^2-y^2}\approx n_{2}-n_{3}$ according to Eq. (\ref{eqn:ox2y2}),
the order parameter $O_{x^2-y^2}>0$ ($<0$) corresponds
the orbital polarized state with $n_{xz}>n_{yz}$ ($n_{xz}<n_{yz}$).
Figure \ref{fig:OO-fig} (b) shows the AFQ order
brought by the divergence of $\chi^Q_{xz}({\bm Q})$.
Although the FQ order in (a) would occur earlier,
we expect the AFQ order in (b) would coexist with the FQ order
when the structure transition is the weak first order.
In fact, the reconstruction of the FSs above $T_{\rm N}$
in detwinned Ba(Fe$_{1-x}$Co$_x$)$_2$As$_2$ \cite{ARPES3}
would indicate the presence of the AFQ order \cite{comment3}.

In Fig. \ref{fig:OO-fig} (c),
we show the correspondence between the quadrupole order 
and the $d$-wavefunction with larger electron occupancy.
In the $O_{x^2-y^2}$-type quadrupole order,
the electrons mainly occupy the state $|xz\rangle$ for $O_{x^2-y^2}>0$,
or the state $|yz\rangle$ for $O_{x^2-y^2}<0$.
In the $O_{xz}$-type quadrupole order,
the electrons mainly occupy the state 
$|xy\rangle+|xz\rangle$ for $O_{xz}>0$
($|xy\rangle-|xz\rangle$ for $O_{xz}<0$).


%
Finally, we make comparison between the present study
and the previous work based on the RPA \cite{Ohno2},
which claims that the divergence of Eq. (\ref{eqn:Cx66})
is caused by large $g_{66}a_{66}\lesssim1$ while neglecting $\chi_{66}^{\rm TO}$.
However, the obtained $O_{x^2-y^2}$-quadrupole order is ``incommensurate''
 \cite{comment5}.
This result highlights the importance of the two-orbiton process 
$\chi_{66}^{\rm TO}$ in order to produce the 
``$\q=0$'' orthorhombic structure transition.


\subsection{Stripe magnetic order produced by 
$O_{x^2-y^2}$-FQ order}

In under-doped iron pnictides,
the collinear-SDW order is induced in the orthorhombic phase
at $T_{\rm N}$, which is slightly lower temperature than $T_S$.
Various explanations for the origin of this SDW transition 
had been proposed previously.
From a strong-coupling scheme,
square-lattice Heisenberg model with in-plane anisotropy,
$J_{1a}$-$J_{1b}$-$J_{2}$ model, had been studied \cite{frustration}.
According to the neutron scattering on CaFe$_2$As$_2$ \cite{Dai},
the high-energy spin-wave dispersion indicates the relation
$J_{1a}\gg J_{1b}$ in the orthorhombic phase.
In this case, experimentally observed staggered spin order 
along the $x$-axis ($a$-axis) is expected to be realized.
However, such strong in-plane anisotropy $J_{1a}\gg J_{1b}$ 
is surprising, considering the small orthorhombicity 
$(a-b)/(a+b)\sim0.003$.
These fact would indicate the existence of orbital or quadrupole order 
in the orthorhombic phase.

\begin{figure}[!htb]
\includegraphics[width=0.98\linewidth]{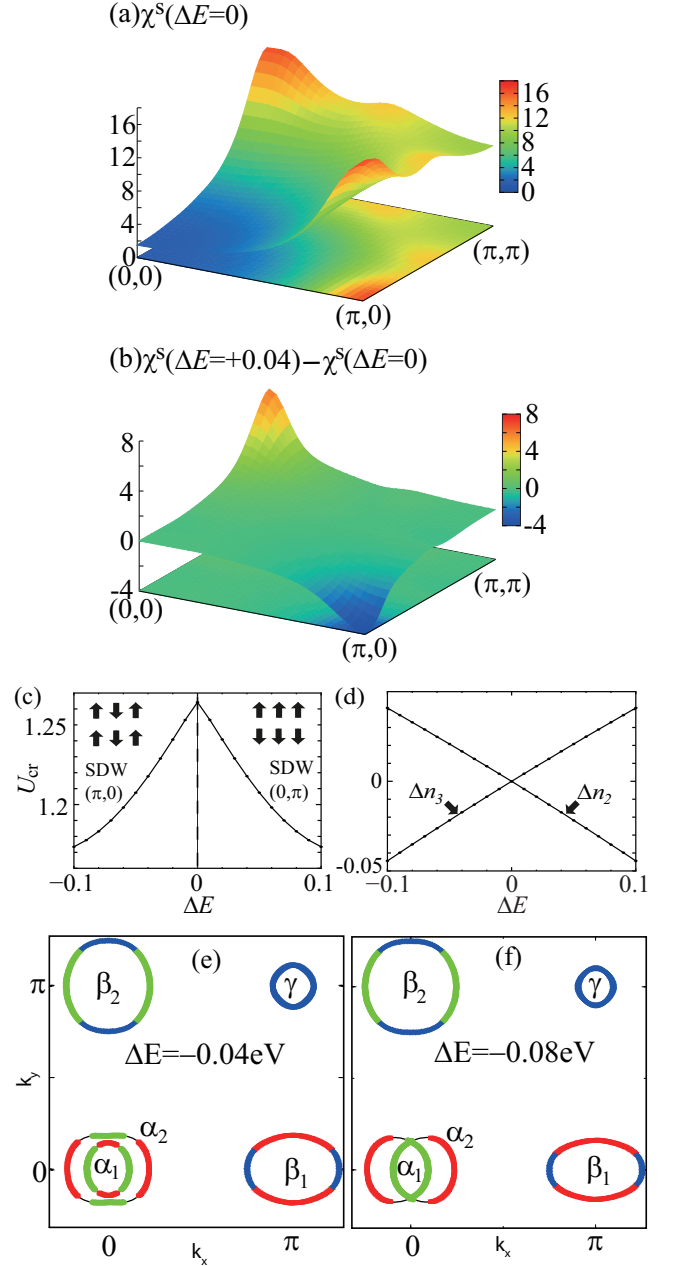}
\caption{
(Color online) 
(a) $\chi^s(\q;\Delta E)$ for $\Delta E=0$.
Used model parameters are $n=6.05$, $U=1.1$, $g=0$, and $T=0.05$
(b) $\chi^s(\q;\Delta E)-\chi^s(\q;0)$ for $\Delta E=0.04$.
Model parameters are the same as those in (a).
(c) $U_c$ as function of $\Delta E$ given by the RPA for $m^*/m=1$.
(d) $\Delta n_{2}$ and $\Delta n_{3}
$ as function of $\Delta E$ for $m^*/m=1$.
Note that $\Delta n\equiv \Delta n_{2}- \Delta n_{3}$.
(e) FSs for $\Delta E=-0.04$ for $n=6.05$. 
We use the same color coding as in Fig. \ref{fig:fig1}.
(f) FSs for $\Delta E=-0.08$ for $n=6.05$. 
In both cases, the best nesting vector is $\q=(\pi,0)$.
}
\label{fig:sdw}
\end{figure}

In this subsection, we study the origin of SDW state
based on the weak-coupling approach.
Hereafter, we assume that $x$-axis corresponds to $a$-axis
(longer lattice constant) in the orthorhombic phase.
In the previous subsection,
we explained that the two-orbital process induces
the $O_{x^2-y^2}$-FQ order in Fig. \ref{fig:OO-fig}.
Note that $O_{x^2-y^2}\approx n_{2}-n_{3}$ according to Eq. (\ref{eqn:ox2y2}).
The corresponding mean-field is given as
%
\begin{eqnarray}
 H'=\Delta E \sum_i \left( |2\rangle\langle 2|
-|3\rangle\langle 3| \right)_i,
\end{eqnarray}
which raises (lowers) the energy-level of orbital 2 (3) by $\Delta E$.
In a similar model, the change in the DOS and FSs by the 
orthorhombic potential $\Delta E$ was studied 
by Chen {\it et al.} \cite{Chen}.
Here, we study the change in the spin susceptibility 
by $\Delta E$ using the RPA.

We calculate the total spin susceptibility
$\chi^s(\q,0)=\sum_{l,m}\chi_{l,l;m,m}^s(\q,0)$ 
for $U=1.1$ and $g=0$.
Figure \ref{fig:sdw} (a) shows the obtained $\chi^s(\q,0)$
for $\Delta E=0$; the corresponding spin Stoner factor is $\a_S=0.87$.
When $\Delta E$ is finite, the four-hold symmetry in $\chi^s(\q,0)$ 
disappears quickly.
Figure \ref{fig:sdw} (b) shows the change in the spin susceptibility,
$\chi^s(\q;\Delta E)-\chi^s(\q;0)$, induced by $\Delta E=+0.04$.
We see that $\chi^s(\q,0)$ increases by $+6.5$ at $\q=(0,\pi)$
while decreases by $-4.0$ at $\q=(\pi,0)$.
Therefore, magnetic frustration is resolved 
and stripe-SDW order can be induced by small $\Delta E$.

Figure \ref{fig:sdw} (c) shows the $\Delta E$-$U_c$ phase diagram
given by the RPA, that is, by the mean-field-approximation.
It is noteworthy that $U_c$ quickly decreases in proportion to 
$|\Delta E|$, because of the degeneracy of orbital 2 and 3.
When $U\lesssim U_{\rm c}$, the experimental SDW order 
with momentum $\q=(\pi,0)$ is realized by the negative 
potential $\Delta E$ that corresponds to $n_2>n_3$.
Figure \ref{fig:sdw} (d) gives the relation between 
$\Delta n=\Delta n_{2}-\Delta n_{3}$ and $\Delta E$:
If we take the band-renormalization effect into account,
we obtain the relation $\Delta n= -0.85(m^*/m)\Delta E$.
According to (c) and (d), we obtain the reduction in $U_c$
due to the FQ order is
\begin{eqnarray}
\Delta U_c= -1.4|\Delta n|= -1.2(m^*/m)|\Delta E|.
 \label{eqn:UcDnDe}
\end{eqnarray}
Therefore, only few percent $\Delta n$ can induce large change 
in $U_{\rm c}$ that is {\it linear in $|\Delta n|$}.
According to recent ARPES measurement in detwinned BaFe$_2$As$_2$ 
\cite{ARPES3}, $\Delta E\sim-0.03$eV in the orthorhombic phase,
which corresponds to $\Delta n=+0.026(m^*/m)$ 
and $\Delta U_c=-0.036(m^*/m)$ in the present five-orbital model.
In this case, the realized SDW order is $\q=(\pi,0)$,
which is consistent with famous strip-type SDW state
in mother compounds \cite{review}.

We can show that the SDW temperature $T_{\rm N}$ also increases
{\it linearly in $|\Delta n|$} based on the Landau theory.
The free energy in the present problem would be given as
\begin{eqnarray}
F(\Delta n)=F(0)+ c\Delta n(m_{(\pi,0)}^2-m_{(0,\pi)}^2),
\end{eqnarray}
where $m_{\Q}$ is the AF order with momentum $\Q$, and 
$F(0)=a\cdot (T-T_{\rm N}^0)(m_{(\pi,0)}^2+m_{(0,\pi)}^2)
+b(m_{(\pi,0)}^4+m_{(0,\pi)}^4)/2 +\cdots$ with $a,b>0$.
Then, we obtain $T_{\rm N}=T_{\rm N}^0+ |c\Delta n|/a$.
The present study shows that $c<0$,
which seems consistent with the numerical result 
in Ref. \onlinecite{OO-theory-MF2}.

Now, we consider the reason why SDW order is produced by $\Delta E$:
Figure \ref{fig:sdw} (e) and (f) shows the change in the 
the FS structure with $\Delta E$.
We can recognize that the intra-orbital (orbital 3) nesting 
between FS $\a_2$ and FS $\b_1$ becomes better, compared to the 
case of $\Delta E=0$ in Fig. \ref{fig:fig1} (a).
Therefore, the origin of the ``FQ-order-induced stripe-SDW'' is 
the ``anisotropy in the intra-orbital nesting''
caused by small $|\Delta E|\sim 0.03$eV,
which corresponds to 
a small orbital polarization $|\Delta n|\sim0.026(m^*/m)$.
This result is consistent with the very small orthorhombicity 
$(a-b)/(a+b)\lesssim0.003$ in the orthorhombic state \cite{review}.
In the strong-coupling description,
the origin of the stripe-SDW state is the 
in-plane anisotropy in the exchange interaction ($J_{1a}\ne J_{1b}$)
 \cite{frustration,Dai} brought by two-orbiton process.

If we go beyond the RPA,
the SDW state will be further stabilized by 
the reduction in the quasiparticle damping $\gamma$
when the FQ-order is established \cite{Onari}:
In fact, in the FLEX approximation \cite{Onari}, $\chi^s(\q)$ is 
suppressed by $\gamma$ due to strong orbital fluctuations in the normal state.
Since the orbital fluctuations is suppressed
when the AFQ-order sets in,
the resultant increment in $\chi^s(\q)$ would stabilize the SDW phase.


\begin{figure}[!htb]
\includegraphics[width=0.95\linewidth]{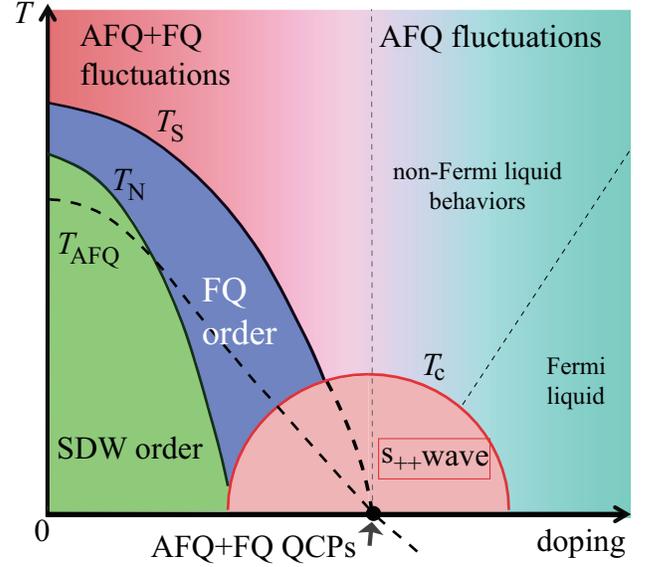}
\caption{
(Color online) 
The phase-diagram for iron-pnictide superconductors
obtained by the present orbital fluctuation theory.
$T_S$ is the orthorhombic transition temperature 
(= FQ order temperature), and $T_{\rm N}$ is SDW transition temperature.
The fact that two QCPs at $T_S=0$ and $T_{\rm AFQ}=0$ almost coincide
means that novel ``multi orbital QCPs'' are realized in iron pnictides.
The left-hand (right-hand) side of the vertical dotted line 
corresponds to $T_{\rm AFQ}>0$ ($T_{\rm AFQ}<0$), in which the 
two-orbiton process is relevant (irrelevant).
At $T_{\rm AFQ}$, the AFQ-order does not occur 
since it is prevented by the FQ-order at $T_S$.
}
\label{fig:phase}
\end{figure}

\subsection{Summary}
In the present paper,
we have studied the realistic five-orbital HH model for iron pnictides.
In the RPA, only the $O_{xz}$-AFQ fluctuations develop
as shown previously \cite{Kontani}, and therefore the softening of 
shear moduli ($C_{66}$, $C_{44}$ and $C_E$) cannot be reproduced.
In the present study beyond the RPA, we revealed that
both $O_{xz}$-AFQ fluctuations and $O_{x^2-y^2}$-FQ fluctuations 
develop at the same time.
The former and the latter fluctuations are the origins of the 
$s_{++}$-wave superconductivity and the orthorhombic structure 
transition, respectively.
The commensurate FQ fluctuations are brought by the two-orbiton process
in Fig. \ref{fig:AL} (c) that is dropped in the RPA.
[In the mean-field theory, the orbital order due to large $g_{66}$
is always ``incommensurate'' \cite{comment5}.]
Fluctuation-induced softening occurs only in $C_{66}$ out of three shear moduli
because of the orbital selection rule for the three-point vertex.
The origin of softening would be interpreted as ``virtual
anharmonicity of lattice vibrations'' that is induced by AFQ fluctuations;
see Fig.\ref{fig:AL} (b).
Possible quadrupole orders in the ordered state
are show in Fig. \ref{fig:OO-fig}.
Using the two-orbiton term in Eq. (\ref{eqn:chi66-analytic2}),
we can fit the recent experimental data of $C_{66}$ in 
Ba(Fe$_{1-x}$Co$_x$)$_2$As$_2$ \cite{comment4} for wide range of doping, 
only by choosing $T_{\rm AFQ}$ while other parameters are fixed.
This fact is a strong evidence for the success of orbital fluctuation theory
in iron pnictide superconductors.

In addition, 
we should stress that the stripe-type antiferro-magnetic state 
is realized in the orbital-ordered state, 
since the small orbital polarization ($\Delta n\lesssim0.05$) can cause
large in-plane anisotropy in the exchange interaction ($J_{1a}\ne J_{1b}$).
Thus, the present study presents a microscopic justification for
the anisotropic Heisenberg model description
for the SDW state \cite{frustration,Dai}.

In Fig. \ref{fig:phase}, we summarize the phase-diagram
of iron-pnictides given by the present orbital fluctuation theory
beyond the RPA.
We stress that $T_{\rm AFQ}$, which is determined experimentally from $C_{66}$,
is positive in the under-doped case ($T_S>0$) while it is negative
in the over-doped case, as recognized from Eq. (\ref{eqn:TS})
obtained in the classical approximation.
Especially, $T_{S}\approx0$ for $T_{\rm AFQ}=0$,
consistently with experiments \cite{Yoshizawa,Goto}.
This result indicats that QCPs for AFQ and FQ orders almost coincide
at the endpoint of the orthorhombic phase.
The emergence of ``multi orbital QCPs'' is favorable to
the orbital-fluctuation-mediated $s_{++}$-wave SC state
 \cite{Kontani,Saito,Onari}.
In fact, 
$T_{\rm AFQ}$ is derived from experimentally observed $C_{66}$ as follows: 
In under-doped systems with $T_S>0$, $T_{\rm AFQ}$ is given 
the Weiss temperature of $C_{66}\propto (T-T_S)/(T-\theta)$, and 
$T_{\rm AFQ}=\theta$ is indeed positive experimentally. 
In over-doped systems, both $T_S$ and $\theta$ are negative ($T_S>\theta$), 
and $C_{66}$ starts to deviate from the Curie-Weiss behavior. 
These experimental results are the strong evedence for the 
realization of the two-orbiton process ($\chi_{66}^{\rm TO}\sim T/(T-T_{\rm AFQ})$)
in iron-pnictides.
In contrast, in the cooperative Jahn-Teller scenario
due to large $g_{66}$ by Yanagi {\it et al.} \cite{Ohno2}, the parameter 
$\theta$ is always negative; see Appendix \ref{sec:ApB} in detail.
Finally, we note that the two-orbiton term $\chi_{66}^{\rm TO}$
in the present study is very similar to the bare 
nematic susceptibility $\chi_{\rm 0,nem}$ in Ref. \cite{Fernandes},
which is the two-magnon term on different sublattices in our terminology.

In summary, the present study 
can explain the superconductivity, orthorhombic transition,
and softening of $C_{66}$ due to FQ and AFQ quantum-criticalities.
The stripe-SDW order is naturally produced 
by the ``orthorhombicity'' of the FQ order.
These results are strong evidence for the realization of the 
orbital-fluctuation-mediated $s_{++}$-wave superconductivity in iron pnictides.
Finally, we stress that the present study enables us to derive the 
important parameters in the orbital fluctuation model in 
Eqs. (\ref{eqn:chi-analytic})-(\ref{eqn:w0})
from the experimental data of shear modulus.




\acknowledgements
We are grateful to M. Yoshizawa for valuable discussions
on his interesting experimental results.
We also thank D.S. Hirashima, M. Sato, Y. Matsuda, T. Shibauchi,
and R.M. Fernandes for valuable comments and discussions.
This study has been supported by Grants-in-Aid for Scientific 
Research from MEXT of Japan, and by JST, TRIP.
Numerical calculations were performed using the facilities of 
the supercomputer center, Institute for Molecular Science.

\appendix
\section{Structure transitions in itinerant electron systems:
Quadrupole-quadrupole interaction V.S. Cooperative Jahn-Teller effect
\label{sec:ApB}}

In this paper, we studied the structure transition 
due to $O_{x^2-y^2}$-FQ order in iron-pnictides.
The ferro quadrupole interaction originates from the 
two-orbiton process with respect to $O_{xz}$-AFQ fluctuations,
which are induced by optical phonons.
The $O_{x^2-y^2}$-FQ fluctuations give the 
softening of the elastic constant $C_{66}$.

In this section, we present a general theory
for the structure transitions in itenerant metals,
and consider the uniqueness of iron pnictides in the next step.
The structure transition due to ferro-quadrupole order
is classified as the ``cooperative Jahn-Teller type''
or ``quadrupole-quadrupole interaction type'' \cite{Levy}.
The elastic constant $C_\phi$ is given by
Eqs. (\ref{eqn:Cphi}) and (\ref{eqn:Cx2}),
where $\chi_\phi$ is the quadrupole susceptibility at $\q=0$
without acoustic phonons.
Now, we introduce the quadrupole-quadrupole interaction $g$
like in Eq. (\ref{eqn:Hint}),
the origin of which is the optical phonons or Coulomb interaction.
In the RPA, $\chi_\phi=\chi_{\phi}^0/(1-g\chi_{\phi}^0)$,
where $\chi_{\phi}^0$ is the bare quadrupole susceptibility.
Then, $C_\phi$ is given as
\begin{eqnarray}
\frac{C_\phi}{C_{\phi,0}}=
\frac{1-(g_{\rm ac}+g)\chi_\phi^0}{1-g\chi_\phi^0}
 \label{eqn:C0C}
\end{eqnarray}
In rare-earth metals in which $f$-electrons are localized,
$\chi_\phi^0$ is proportional to $1/T$ \cite{Levy,Thal}.
However, this replacement is inappropriate in itinerant metals.
In nearly ferro-quadrupole itinerant metals,
$|1-(g_{\rm ac}+g)\chi_\phi^0|\ll1$ at zero temperature.

Here, we consider the temperature dependence of $C_\phi$ beyond the RPA.
In the FLEX approximation \cite{Onari} or SCR theory \cite{Moriya},
$\chi_\phi^0$ is replaced with $\chi_\phi^0-\a T$ ($\a>0$)
due to the thermal fluctuations, which are mainly described as the self-energy.
In this case, Eq. (\ref{eqn:C0C}) becomes
\begin{eqnarray}
\frac{C_\phi}{C_{\phi,0}}\approx \frac{T-T_S}{T-\theta}
 \label{eqn:ApB2}
\end{eqnarray}
where $T_S=-(1-(g_{\rm ac}+g)\chi_\phi^0)/(g_{\rm ac}+g)\a$
and $\theta=-(1-g\chi_\phi^0)/g\a$.
Therefore, $C_\phi$ shows the Curie-Weiss behavior in the RPA
by taking the thermal fluctuations into account.

First, we consider the case (i) $g_{\rm ac}\ll g$, in which
the structure transition is driven by quadrupole-quadrupole interaction
 \cite{Levy}.
In this case, $E_{\rm JT}\equiv T_S-\theta \approx g_{\rm ac}/g^2\a$,
which is much smaller than $T_S$.
(Note that $C_\phi$ is not soften when $g_{\rm ac}=0$.)
Then, the lattice distortion in the ordered state 
will be very small ($\ll 1$\%), like in PrRu$_4$P$_{12}$. 
In the opposite case (ii) $g_{\rm ac}\gg g$,
the structure transition is driven by cooperative Jahn-Teller effect
 \cite{Levy}.
In this case, $\theta\approx -1/g\a$, which takes a large negative value
since $\a$ is small; $\theta\sim -300$K.
Then, the lattice distortion in the ordered state 
should be very large ($\gg 1$\%), like manganites.
In case (ii), the energy of the ``ferro-orbital fluctuations'' 
induced by the acoustic phonon is very low, and the $T_{\rm c}$ of
the orbital-fluctuation superconductivity is still lower.
Therefore, emergence of high-$T_{\rm c}$ superconductivity is 
unlikely in case (ii).

In the case of $C_{66}$ in iron pnictides,
$O_{x^2-y^2}$-$O_{x^2-y^2}$ interaction $g$ due to optical phonons is absent.
Moreover, $O_{x^2-y^2}$-FQ fluctuations due to Coulomb interaction
do not develop even in the $U'>U$ model \cite{Ohno1}.
Therefore, if we try to explain the $C_{66}$ softening within the RPA,
we have to assume the case (ii), {\it i.e.}, the cooperative Jahn-Teller 
transitionas, as proposed by Yanagi {\it et al.} \cite{Ohno2}.
However, it contradicts with experiments facts, that is,
$(a-b)/(a+b)\lesssim0.3$\% and $n_{xz}-n_{yz}\lesssim 5$\%
in the orthorhombic phase irrespective of higher $T_S$ ($\sim100$K),
and $\theta>0$ in the under-doped pnictides.
Therefore, the RPA analysis done by Yanagi {\it et al.} \cite{Ohno2}
is inconsistent with experiments.

In this paper, we studied the structure transition in case (i), {\it i.e.},
the quadrupole-quadrupole interaction type.
The origin of quadrupole-quadrupole interaction is the 
``two-orbiton process'', which is not taken into account in the RPA.
In two-dimensional systems,
the two-orbiton process gives the Curie-Weiss behavior 
in Eq. (\ref{eqn:ApB2}), if allowed by the orbital selectrion rule
discussed in Sec. \ref{sec:Dis-C66}.
By using reasonable sets of parameters ($g_{\rm ac}<g$), 
experimentally observed Curie-Weiss behavior of $C_{66}$ 
in under-doped pnictides ($T_S>0$) \cite{Yoshizawa,Goto}
is well reproduced, given in Eq. (\ref{eqn:C66-temp}):
According to Eq. (\ref{eqn:TS}), $T_{\rm AFQ} \ (=\theta)$ is positive in 
under-doped pnictides, and $T_{\rm AFQ}\approx0$ at the critical point $T_S=0$, 
consistently with experimental reports \cite{Goto}.
In over-doped pnictides, the softening of $C_{66}$
becomes moderate since the two-orbiton process gives
very weak temperature dependence.

\section{Quadrupole susceptibilities in the ten-orbital model:
effect of unfold gauge transformation
\label{sec:10}}

\begin{figure}[!htb]
\includegraphics[width=0.99\linewidth]{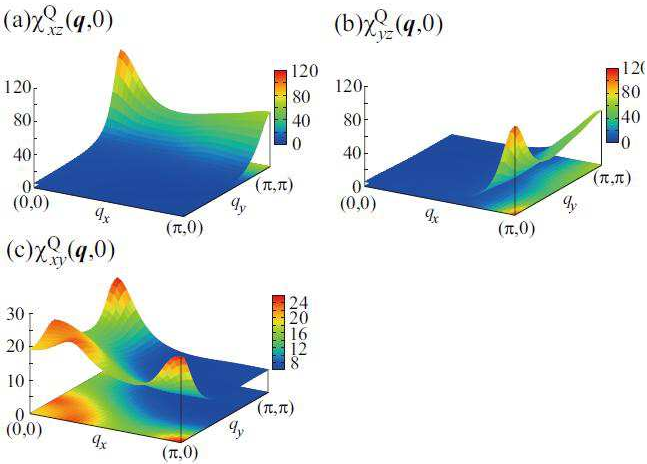}
\caption{
(Color online) 
Quadrupole susceptibilities in the ten-orbital model for
(a) $\chi_{xz}^{Q}(\q)$, (b) $\chi_{yz}^{Q}(\q)$, and 
(c) $\chi_{xy}^{Q}(\q)$, respectively.
We put $n=6.05$, $T=0.05$, and $\a_c=0.98$.
}
\label{fig:chic10}
\end{figure}

\begin{figure}[!htb]
\includegraphics[width=0.99\linewidth]{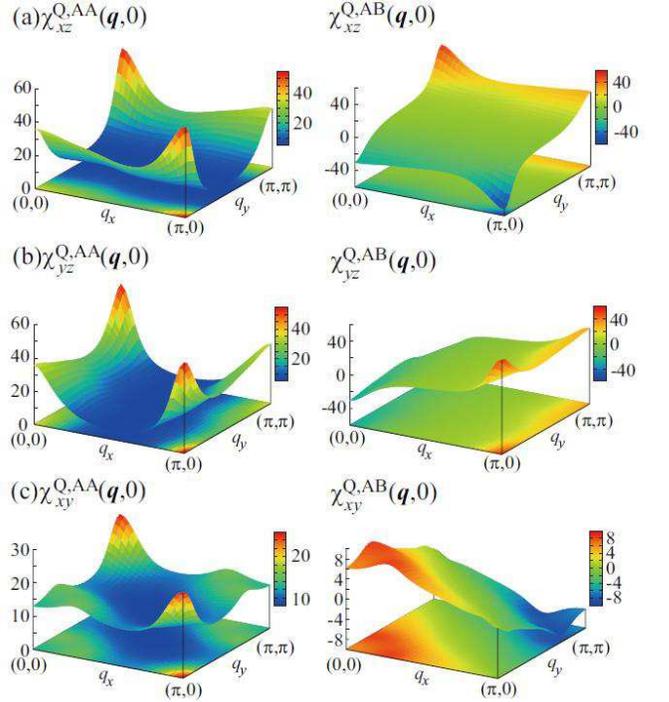}
\caption{
(Color online) 
Quadrupole susceptibilities in the ten-orbital model for 
(a) $\chi_{xz}^{Q,AA}(\q)$ and $\chi_{xz}^{Q,AB}(\q)$, 
(b) $\chi_{yz}^{Q,AA}(\q)$ and $\chi_{yz}^{Q,AB}(\q)$, and 
(c) $\chi_{xy}^{Q,AA}(\q)$ and $\chi_{xy}^{Q,AB}(\q)$,
respectively.
We put $n=6.05$, $T=0.05$, and $\a_c=0.98$.
}
\label{fig:chic10-2}
\end{figure}

In Sec. \ref{sec:RPA}, 
we have calculated $\chi_\Gamma^Q(\q,0)$ based on the five-orbital model
using the RPA, shown in Fig. \ref{fig:chic}.
In this Appendix, we discuss the effect of the unfold-gauge transformation
in deriving the five-orbital model on $\chi_\Gamma^Q(\q,0)$.
This gauge transition changes the signs of $2,3$-orbitals for Fe-B sites. 
Therefore, signs of quadrupole operators ${\hat O}_{xz/yz}^i$
at Fe-B sites are reversed, as recognized in 
Eqs. (\ref{eqn:oxz})-(\ref{eqn:oz2}).
For this reason, $\chi_{xz,yz}^Q(\q,\w)$ is not gauge-invariant since
it contains linear terms with respect to ${\hat O}_{xz/yz}^i$,
although the el-el interaction in 
Eqs. (\ref{eqn:Gamma-s})-(\ref{eqn:Gamma-c}) is gauge invariant.

From now on, we calculate the quadrupole susceptibilities
in the ten-orbital model.
We describe the orbitals of Fe-A (Fe-B) ions as $1\sim5$ ($6\sim10$).
In the RPA, the susceptibilities are given as
\begin{eqnarray}
\chi_{\Gamma}^Q(q)&=&\chi_{\Gamma}^{Q,AA}(q)+\chi_{\Gamma}^{Q,AB}(q),
 \label{eqn:chiQ10} \\
\chi_{\Gamma}^{Q,AA(AB)}(q)&=&
\sum_{ll'}^{A}\sum_{mm'}^{A(B)}o_\Gamma^{ll'}\chi_{ll'mm'}^c(q)o_\Gamma^{mm'},
 \label{eqn:chiQ-AA} 
\end{eqnarray}
for $\Gamma=xz,yz,xy$,
where $o_\Gamma^{lm}=o_\Gamma^{l-5,m-5}$ for $l,m\ge6$.
The obtained quadrupole susceptibilities in Eq. (\ref{eqn:chiQ10})
is show in Fig. \ref{fig:chic10}, and its diagonal 
and off-diagonal terms with respect to A and B are also shown 
in Fig. \ref{fig:chic10-2}.
In these figures, we have unfolded the susceptibilities into the 
single-iron BZ in order to make comparison with Fig. \ref{fig:chic}.
According to Eq. (\ref{eqn:chiQ10}),
$\chi_\Gamma^{Q}(q)$ in the ten-orbital model 
is given by $\chi_{\Gamma}^{Q,AA}(q)+\chi_{\Gamma}^{Q,AB}(q)$.
We see that $\chi_{xz}^Q(\q,0)$ is Fig. \ref{fig:chic10} (a) has
a sharp peak at $\q=(0,\pi)$, since both $\chi_{xz}^{Q,AA}(q)$
and $\chi_{xz}^{Q,AB}(q)$ in Fig. \ref{fig:chic10-2} (a)
have peaks at $\q=(0,\pi)$.

If we perform the unfold-gauge transformation,
the sign of $\chi_{\Gamma}^{Q,AB}(q)$ is inverted for $\Gamma=xz$ and $yz$.
Therefore, in the five-orbital model,
$\chi_\Gamma^{Q}(q,0)=\chi_{\Gamma}^{Q,AA}(q)-\chi_{\Gamma}^{Q,AB}(q)$
for $\Gamma=xz$ and $yz$.
For this reason, the peak in $\chi_{xz}^Q(\q,0)$
move from $\q=(0,\pi)$ to $(\pi,0)$ under the gauge transformation,
consistently with the result in Fig. \ref{fig:chic} (a).
In contrast to $\chi_{xz/yz}^{Q,AB}(\q)$, $\chi_{xy}^{Q}(\q)$ is gauge invariant.
For this reason, Fig. \ref{fig:chic10} (c) in the ten-orbital model
coincides with Fig. \ref{fig:chic} (c) in the five-orbital model.



\begin{thebibliography}{99}

\bibitem{Hosono}
Y. Kamihara, T. Watanabe, M. Hirano, and H. Hosono:
J. Am. Chem. Soc. {\bf 130}, 3296 (2008).

\bibitem{Kuroki}
K. Kuroki, S. Onari, R. Arita, H. Usui, Y. Tanaka, H. Kontani, and H. Aoki,
Phys. Rev. Lett. {\bf 101}, 087004 (2008).

\bibitem{Mazin}
I. I. Mazin, D. J. Singh, M. D. Johannes, and M. H. Du,
Phys. Rev. Lett. {\bf 101}, 057003 (2008).

\bibitem{hirschfeld}
S. Graser, G. R. Boyd, C. Cao, H.-P. Cheng, P. J. Hirschfeld, 
and D. J. Scalapino, Phys. Rev. B {\bf 77}, 180514(R) (2008).

\bibitem{chubukov}
A. V. Chubukov, D. V. Efremov, and I. Eremin,
Phys. Rev. B {\bf 78}, 134512 (2008). 

\bibitem{Onari-impurity}
S. Onari and H. Kontani, 
Phys. Rev. Lett. {\bf 103} 177001 (2009).

\bibitem{Sato-imp}
A. Kawabata, S. C. Lee, T. Moyoshi, Y. Kobayashi, and M. Sato,
J. Phys. Soc. Jpn. {\bf 77} (2008) Suppl. C 103704;
M. Sato, Y. Kobayashi, S. C. Lee, H. Takahashi, E. Satomi, and Y. Miura,
J. Phys. Soc. Jpn. {\bf 79} (2009) 014710;
S. C. Lee, E. Satomi, Y. Kobayashi, and M. Sato,
J. Phys. Soc. Jpn. {\bf 79} (2010) 023702.

\bibitem{Nakajima}
Y. Nakajima, T. Taen, Y. Tsuchiya, T. Tamegai, H. Kitamura, and T. Murakami,
Phys. Rev. B {\bf 82}, 220504 (2010).

\bibitem{Kontani}
H. Kontani and S. Onari, Phys. Rev. Lett. {\bf 104}, 157001 (2010).

\bibitem{Saito}
T. Saito, S. Onari, and H. Kontani,
Phys. Rev. B {\bf 82}, 144510 (2010).

\bibitem{Onari}
S. Onari and H. Kontani, arXiv:1009.3882

\bibitem{Lee}
C.-H. Lee, A. Iyo, H. Eisaki, H. Kito, M. T. Fernandez-Diaz, 
T. Ito, K. Kihou, H. Matsuhata, M. Braden, and K. Yamada,
J. Phys. Soc. Jpn. {\bf 77}, 083704 (2008).

\bibitem{Shimo-Science}
T. Shimojima, F. Sakaguchi, K. Ishizaka, Y. Ishida, T. Kiss, M. Okawa, 
T. Togashi, C.-T. Chen, S. Watanabe, M. Arita, K. Shimada, H. Namatame, 
M. Taniguchi, K. Ohgushi, S. Kasahara, T. Terashima, T. Shibauchi, 
Y. Matsuda, A. Chainani, and S. Shin,
Sccience {\bf 332}, 564 (2011).

\bibitem{res-exp}
A. D. Christianson, E. A. Goremychkin, R. Osborn, S. Rosenkranz, M. D. Lumsden,
C. D. Malliakas, I. S. Todorov, H. Claus, D. Y. Chung, M. G. Kanatzidis,
R. I. Bewley, and T. Guidi,
Nature {\bf 456}, 930 (2008);
Y. Qiu, W. Bao, Y. Zhao, C. Broholm, V. Stanev, Z. Tesanovic, Y. C. Gasparovic,
S. Chang, J. Hu, B. Qian, M. Fang, and Z. Mao,
Phys. Rev. Lett. {\bf 103}, 067008 (2009);
D. S. Inosov, J. T. Park, P. Bourges, D. L. Sun, Y. Sidis, A. Schneidewind, K. Hradil,
D. Haug, C. T. Lin, B. Keimer, and V. Hinkov,
Nature Physics {\bf 6}, 178 (2010).

\bibitem{res-the1}
T. A. Maier and D. J. Scalapino, Phys. Rev. B {\bf 78}, 020514(R) (2008);
T. A. Maier, S. Graser, D.J. Scalapino, and P. Hirschfeld, 
Phys. Rev. B {\bf 79}, 224510 (2009).
\bibitem{res-the2}
M. M. Korshunov and I. Eremin, Phys. Rev. B {\bf 78}, 140509(R) (2008).

\bibitem{Onari-resonance}
S. Onari, H. Kontani, and M. Sato,
Phys. Rev. B {\bf 81}, 060504(R) (2010) 

\bibitem{Yasui}
M. Sato {\it et al.}, private communications

\bibitem{Scala}
T.A. Maier, S. Graser, P.J. Hirschfeld, and D.J. Scalapino,
Phys. Rev. B {\bf 83}, 220505(R) (2011).

\bibitem{Matsuda}
S. Kasahara, T. Shibauchi, K. Hashimoto, K. Ikada, S. Tonegawa,
R. Okazaki, H. Shishido, H. Ikeda, H. Takeya, K. Hirata, T. Terashima, and Y. Matsuda,
Phys. Rev. B {\bf 81}, 184519 (2010).

\bibitem{Kontani-review}
 H. Kontani, Rep. Prog. Phys. {\bf 71}, 026501 (2008);
 H. Kontani and K. Yamada, J. Phy. Soc. Jpn. {\bf 74}, 155 (2005);
 H. Kontani, K. Kanki, and K. Ueda, Phys. Rev. B  {\bf 59}, 14723 (1999).

\bibitem{review}
D. C. Jhonston, Advances in Physics {\bf 59}, 803 (2010)

\bibitem{Shimojima}
T. Shimojima, K. Ishizaka, Y. Ishida, N. Katayama, K. Ohgushi, T. Kiss, M. Okawa,
T. Togashi, X.-Y. Wang, C.-T. Chen, S. Watanabe, R. Kadota, T. Oguchi, A. Chainani, and S. Shin,
Phys. Rev. Lett. {\bf 104}, 057002 (2010).

\bibitem{ARPES2}
Q. Wang, Z. Sun, E. Rotenberg, F. Ronning, E. D. Bauer, H. Lin, R. S. Markiewicz,
M. Lindroos, B. Barbiellini, A. Bansil, D. S. Dessau,
arXiv:1009.0271

\bibitem{ARPES3}
M. Yi, D. H. Lu, J.-H. Chu, J. G. Analytis, A. P. Sorini, A. F. Kemper, S.-K. Mo,
R. G. Moore, M. Hashimoto, W. S. Lee, Z. Hussain, T. P. Devereaux, I. R. Fisher, Z.-X. Shen,
PNAS {\bf 108} 6878.

\bibitem{frustration}
T. Yildirim, Physica C {\bf 469} (2009) 425;
F. Kruger, S. Kumar, J. Zaanen, and J. van den Brink,
Phys. Rev. B {\bf 79}, 054504 (2009);
W. Lv, J. Wu, and P. Phillips,
Phys. Rev. B {\bf 80}, 224506 (2009);
C.-C. Chen, B. Moritz, J. van den Brink, T. P. Devereaux, and R. R. P. Singh,
Phys. Rev. B {\bf 80}, 180418 (2009).

\bibitem{rho-nematic}
J.-H. Chu, J. G. Analytis, K. D. Greve, P. L. McMahon, Z. Islam, Y. Yamamoto, and I. R. Fisher,
Science {\bf 329}, 824 (2010);
J. J. Ying, X. F. Wang, T. Wu, Z. J. Xiang, R. H. Liu, Y. J. Yan,
A. F. Wang, M. Zhang, G. J. Ye, P. Cheng, J. P. Hu, and X. H. Chen,
arXiv:1012.2731.

\bibitem{Prosorov}
E. C. Blomberg, M. A. Tanatar, A. Kreyssig, N. Ni, A. Thaler,
R. Hu, S. L. Bud'ko, P. C. Canfield, A. I. Goldman, and R. Prozorov,
Phys. Rev. B {\bf 83}, 134505 (2011).

\bibitem{optical}
A. Dusza, A. Lucarelli, F. Pfuner, J.-H. Chu, I.R. Fisher, L. Degiorgi,
arXiv:1007.2543;
M. Nakajima {\it et al.}, unpublished.



\bibitem{Fernandes}
R.M. Fernandes, L. H. VanBebber, S. Bhattacharya, P. Chandra, 
V. Keppens, D. Mandrus, M.A. McGuire, B.C. Sales, A.S. Sefat, 
and J. Schmalian, Phys. Rev. Lett. {\bf 105}, 157003 (2010) 

\bibitem{Yoshizawa}
M. Yoshizawa, R. Kamiya, R. Onodera, Y. Nakanishi, K. Kihou, H. Eisaki, and C. H. Lee,
arXiv:1008.1479.

\bibitem{comment4}
M. Yoshizawa {\it et al.}, private communication.

\bibitem{Goto}
T. Goto, R. Kurihara, K. Araki, K. Mitsumoto, M. Akatsu, Y. Nemoto, 
S. Tatematsu, and M. Sato, J. Phys. Soc. Jpn. {\bf 80}, 073702 (2011); 
T. Goto {\it et al}. (unpublised)

\bibitem{Dai}
J. Zhao, D. T. Adroja, D.-X. Yao, R. Bewley, S. Li, X. F. Wang, G. Wu, X. H. Chen, J. Hu, and P. Dai,
Nature Physics {\bf 5}, 555 (2009).

\bibitem{Little}
W.A. Little, Phys. Rev. {\bf 134}, A1416 (1964)

\bibitem{Ginzburg}
V.L. Ginzburg, Zh. Eksp. Teor. Fiz. {\bf 47} 2318 (1964)
[Sov. Phys. JETP {\bf 20}, 1549 (1965)]

\bibitem{Hirsch}
J. E. Hirsch and D. J. Scalapino, Phys. Rev. Lett. {\bf 56}, 2732 (1986) 

\bibitem{Tesa}
V. Cvetkovic and Z. Tesanovic, EPL, {\bf 85} (2009) 37002

\bibitem{Sawa}
M. Berciu, I. Elfimov, and G. A. Sawatzky,
arXiv:0811.0214

\bibitem{Takimoto}
T. Takimoto, T. Hotta, T. Maehira and K. Ueda,
J. Phys.: Condens. Matter {\bf 14}, L369 (2002). 

\bibitem{Ohno1}
Y. Yanagi, Y. Yamakawa, and Y. Ono, 
Phys. Rev. B {\bf 81}, 054518 (2010) 

\bibitem{Miyake}
T. Miyake, K. Nakamura, R. Arita, M. Imada, 
J. Phys. Soc. Jpn. {\bf 79}, 044705 (2010)

\bibitem{Ohno2}
Y. Yanagi, Y. Yamakawa, N. Adachi, and Y. Ono,
J. Phys. Soc. Jpn. {\bf 79} (2010) 123707 

\bibitem{comment}
H. Kontani and S. Onari, 
 J. Phys. Soc. Jpn., {\bf 80}, 056001 (2011).

\bibitem{Schrieffer}
J.R. Schrieffer, {\it theory of superconductivity}
(Benjamin, New York, 1964)

\bibitem{comment5}
In Ref. \onlinecite{Ohno2}, $\chi^Q_{x^2-y^2}(\q,0)$ due to the 
``othrorhombic phonon'' ($=g_{66}$ in the present paper) 
is incommensurate with momentum $\q\approx(0.35\pi,0)$.
In the present tight-binding model \cite{Kuroki}, 
$\chi^Q_{x^2-y^2}(\q,0)$ due to $g_{66}$ has the highest peak at 
$\q\approx(\pi,0)$ due to the nesting between electron- and hole-pockets,
while it is hardly enhanced at $\q={\bm0}$.

\bibitem{Yada}
K. Yada and H. Kontani, Phys. Rev. B {\bf 77}, 184521 (2008);
K. Yada and H. Kontani, J. Phys. Soc. Jpn. {\bf 75}, 033705 (2006) 

\bibitem{comment2}
In Ref. \onlinecite{Saito},
we claimed that the enhancement of
$\chi_{xz}^Q({\bm 0},0)$ in the five-orbital model
induces the softening of $C_{44}$.
However, it was incorrect since the effect of unfold-gauge transformation 
on $\chi_{xz}^Q({\bm 0},0)$ was overlooked.
In fact, $\chi_{xz}^Q({\bm 0},0)$ is not enhanced in the ten-orbital model.

\bibitem{Kontani-VV}
H. Kontani, and K. Yamada, J. Phys. Soc. Jpn. {\bf 65}, 172 (1996)

\bibitem{Boeri}
L. Boeri, O. V. Dolgov, and A. A. Golubov,
Phys. Rev. Lett. {\bf 101}, 026403 (2008) 

\bibitem{Thal}
P. Thalmeier and B. Luthi, in {\it Handbook on the Physics
and Chemistry of Rare Earths} Vol. 14, ed K A Gschneidner
Jr and L Eyring (Amsterdam: Elsevier) p 245 (1991). 

\bibitem{Levy}
P. M. Levy, P. Morin and D. Schmitt, Phys. Rev. Lett. {\bf 42}, 1417 (1979)

\bibitem{AL}
L. G. Aslamasov and A. I. Larkin, 
Soviet Phys.--SolidState {\bf 10}, 875 (1968).

\bibitem{Fuku}
H. Fukuyama, H. Ebisawa, and T. Tsuzuki,
Prog. Theor. Phys. {\bf 46} (1971) 1028.

\bibitem{Uss}
I. Ussishkin, Phys. Rev. B {\bf 68}, 024517 (2003)

\bibitem{Moriya}
T. Moriya, {\it Spin Fluctuations in Itinerant Electron Magnetism}
(Springer-Verlag, 1985);
T. Moriya and K. Ueda, Adv. Physics {\bf 49} 555 (2000);
T. Moriya and K. Ueda : Rep. Prog. Phys. {\bf 66} (2003) 1299.

%

\bibitem{AGD} A.A. Abricosov, L.P. Gor'kov and I.E. Dzyaloshinskii,
 {\it Methods of Quantum Field Theory in Statistical Physics}
 (Dover, New York, 1975).

\bibitem{Shishido}
H. Shishido, A. F. Bangura, A. I. Coldea, S. Tonegawa, K. Hashimoto, S. Kasahara,
P. M. C. Rourke, H. Ikeda, T. Terashima, R. Settai, Y. Onuki, D. Vignolles,
C. Proust, B. Vignolle, A. McCollam, Y.Matsuda, T. Shibauchi, A. Carrington,
Phys. Rev. Lett. {\bf 104}, 057008 (2010).

\bibitem{OO-theory-MF}
K. Kubo and P. Thalmeier, J. Phys. Soc. Jpn. 78, 083704 (2009);
C.-C. Lee, W.-G. Yin, and W. Ku, Phys. Rev. Lett. {\bf 103}, 267001 (2009);
H. Oh, D. Shin, and H. J. Choi, arXiv:1012.2224.

\bibitem{OO-theory-MF2}
K. Sugimoto, E. Kaneshita, and T. Tohyama, 
J. Phys. Soc. Jpn. {\bf 80} (2011) 033706.

\bibitem{comment3}
H. Kontani, unpublished.

\bibitem{Chen}
C.-C. Chen, J. Maciejko, A. P. Sorini, B. Moritz, R. R. P. Singh, and T. P. Devereaux,
Phys. Rev. B {\bf 82}, 100504(R) (2010).

\end{thebibliography}
\end{document}